# A GN-model closed-form formula considering coherency terms in the Link function and covering all possible islands in 2-D GN integration

**Mahdi Ranjbarzefreh, Pierluigi Poggiolini**

*Politecnico di Torino*

**Abstract:** Efficient evaluation of nonlinearity in modern coherent optical telecommunication networks is necessary for design, optimization and management. The original GN model formula can not be used in real time applications due to long processing time needed for numerical 2-D integration, particularly when facing with full C band or C+L band. There are closed form approximations of the GN model which has been mathematically derived based on some assumptions which fail in low dispersion regime. In this work, we derive a GN closed form formula without any restrictive assumption capable of handling *arbitrary* WDM combs and *arbitrary* link structures in closed form which can be used with a very good accuracy with respect to original GN formula for networks employing both high and low dispersion fibers.

## 1. Introduction

Recently, real-time management and control (M&C) of optical WDM networks has become a very active topic of investigation. A necessary tool, needed to enable real-time M&C, is a computationally fast model of the non-linear propagation disturbance (or NLI, non-linear interference) incurred by the WDM signal, due to the fiber Kerr effect. Many NLI models are currently available. However, all of them, in their original form, contain integrals that need to be evaluated numerically. This makes real-time use of these models highly problematic.

One way to overcome this hurdle is trying to obtain, through suitable assumptions and approximations, *closed-form* formulas for the NLI model, i.e., formulas containing neither numerical integrals nor infinite or otherwise excessively large summations. One early result in this direction, based on the GN model of NLI, is reported in [1]. While effective, [1] did not support fiber dispersion slope vs. frequency, frequency-dependent fiber loss, and inter-channel stimulated Raman scattering. These three features are critical for the modeling of modern ultra-broadband systems, especially those encompassing both C and L band.

Quite recently, two efforts [2] and [3] were carried out independently to generalize the formula [1] and include all the missing features above. Both were successful, leading to similar, though not identical, equations. Despite the substantial progress that they represent, these models still have limitations.

One limitation is that both [2] and [3], as well as [1] from which they derive, are based on the *incoherent* GN model, or iGN model. The iGN model, as opposed to the GN model, neglects the *coherent* interference between the NLI generated by different spans in an optical fiber link. In fact, the iGN model assumes that the contribution of NLI due to each span is generated independently and is added to the contribution of all other spans in terms of *power*.

Another limitation common to [1]-[3] is the fact that only SCI and XCI are accounted for, but not MCI. SCI, self-channel interference, is the NLI produced by the channel-under-test (CUT) onto itself. XCI, cross-channel interference, is NLI produced by any single one of the channels onto the CUT. Finally, the neglected one, multi-channel interference, is the NLI produced by two or three other channels onto the CUT.

Both incoherent NLI accumulation assumption and neglecting MCI greatly simplify the analytical derivation of a closed-form formula, but they also affect accuracy. In particular, in links which include fiber segments where the dispersion values $|\beta_2|$ is lower than about 1.5 ps$^2$/km, the coherence effect of NLI accumulation occurs over rapidly increasing spectral bandwidths, growing in strength, and some of the MCI terms are no longer negligible. As a result, these approximations cause substantial NLI estimation error.

The goal of this paper is to remove both these approximations while still obtaining a closed-form set of formulas allowing real-time NLI estimation even in links comprising low-dispersion fibers, and even fibers that have an in-band dispersion zero. To reach this goal, we start from the original GN-model formula and from [3]. We then derive a new closed-form GN model formula making neither the incoherent accumulation approximation nor neglecting the MCI contributions.

As a further introductory remark, we point out that another limitation of [1]-[3] is that they try to reproduce the GN model, rather than the EGN model [4]. This essentially means that some loss of accuracy can be expected when using modulation formats employing other than Gaussian-shaped constellations. However, in [5] the formula [3] was corrected using a machine-learning approach, essentially turning [3] into an EGN model closed-form approximation. The same approach can be used for the more general formula presented here, to turn it into a very general closed-form approximation of the EGN model. However, we consider this further step beyond the scope of this paper and it will not be dealt with here.

Finally, while this paper is completely self-contained, we advise the reader that it may be quite helpful to familiarize with the contents of [3] before reading this document.

## 2. GN model formula

The general formula of the GN model is [1,6]:

$$G_{NLI}(f) = \frac{16}{27} \int_{-\infty}^{+\infty} \int_{-\infty}^{+\infty} G_s(f_1) G_S(f_2) G_S(f_1 + f_2 - f) \times |LK(f_1, f_2, f_1 + f_2 - f)|^2 \, df_1 df_2 \tag{1}$$

Where in (1), $G_s(f)$ is the power spectral density (PSD) of the WDM signal launched into the fiber and $LK$ is the link function which is determined based on the fiber link configuration. The WDM PSD can be written in terms of each channel PSD as:

$$G_s(f) = \sum_{m_{ch}=1}^{N_c} G_{m_{ch}}(f) \tag{2}$$

Where $N_c$ is the number of channels available in WDM comb and $G_{m_{ch}}(f)$ is the PSD due to $m_{ch}$'th channel in the WDM comb. Combining (1) and (2) we will have:

| | |
|---|---|
| $$G_{NLI}(f)=\frac{16}{27}\sum_{m_{ch}=1}^{N_c}\sum_{n_{ch}=1}^{N_c}\sum_{k_{ch}=1}^{N_c}\int_{-\infty}^{+\infty}\int_{-\infty}^{+\infty}G_{m_{ch}}(f_1)G_{n_{ch}}(f_2)\times G_{k_{ch}}(f_1+f_2-f)\lvert LK(f_1,f_2,f_1+f_2-f)\rvert^2\,df_1df_2$$ | (3) |

We assume PSD of channels are rectangular shape functions in frequency domain. In general, the PSD of each WDM channel has raised cosine shape while for keeping the simplicity we assume all the channels are rectangular shape. So, we have:

| | |
|---|---|
| $$G_{m_{ch}}(f)=\begin{cases}G_{m_{ch}} & f_{s,m_{ch}}\le f\le f_{e,m_{ch}}\\ 0 & otherwise\end{cases}$$ | (4) |

Where in (4), $f_{s,m_{ch}}$ and $f_{e,m_{ch}}$ are the start and end frequency of the $m_{ch}$'th channel in WDM comb respectively. Also $G_{m_{ch}}$ is the constant value of PSD due to the $m_{ch}$'th channel. So by using equation (4), the equation (3) can be written as:

| | |
|---|---|
| $$G_{NLI}(f)=\frac{16}{27}\sum_{m_{ch}=1}^{N_c}\sum_{n_{ch}=1}^{N_c}\sum_{k_{ch}=1}^{N_c}G_{m_{ch}}G_{n_{ch}}\int_{f_{s,n_{ch}}}^{f_{e,n_{ch}}}\int_{f_{s,m_{ch}}}^{f_{e,m_{ch}}}G_{k_{ch}}(f_1+f_2-f)\lvert LK(f_1,f_2,f_1+f_2-f)\rvert^2\,df_1df_2$$ | (5) |

It is clear from equation (5) that to achieve a closed form formula for the GN model, the 2-D integration must be solved analytically. To do that, we first focus on the 2-D integration area in $f_1-f_2$ plane in section (3) and afterwards, in section (4), we will discuss about finding an integrand function for the function under integral.

## 3. Area of 2-D integration

The $k_{ch}$'th channel, similar to the $m_{ch}$'th channel in equal (4), can be represented as:

| | |
|---|---|
| $$G_{k_{ch}}(f)=\begin{cases}G_{k_{ch}} & f_{s,k_{ch}}\le f\le f_{e,k_{ch}}\\ 0 & otherwise\end{cases}$$ | (6) |

Therefore, we have:

| | |
|---|---|
| $$G_{k_{ch}}(f_1+f_2-f)=\begin{cases}G_{k_{ch}} & f_{s,k_{ch}}\le f_1+f_2-f\le f_{e,k_{ch}}\\ 0 & otherwise\end{cases}=\begin{cases}G_{k_{ch}} & f_{s,k_{ch}}+f\le f_1+f_2\le f_{e,k_{ch}}+f\\ 0 & otherwise\end{cases}$$ | (7) |

Only for notation simplicity, we define:

| | |
|---|---|
| $f'_{s,k_{ch}}\triangleq f_{s,k_{ch}}+f$ | (8) |
| $f'_{e,k_{ch}}\triangleq f_{e,k_{ch}}+f$ | (9) |

So, we have:

| | |
|---|---|
| $G_{k_{ch}}(f_1 + f_2 - f) = \begin{cases} G_{k_{ch}} & f'_{s,k_{ch}} \le f_1 + f_2 \le f'_{e,k_{ch}} \\ 0 & otherwise \end{cases}$ | (10) |

So, the 2-D integration in (5) can be written as:

| | |
|---|---|
| $\int_{f_{s,n_{ch}}}^{f_{e,n_{ch}}} \int_{f_{s,m_{ch}}}^{f_{e,m_{ch}}} G_{k_{ch}}(f_1 + f_2 - f) \|LK(f_1, f_2, f_1 + f_2 - f)\|^2 \, df_1 df_2 = G_{k_{ch}} \times \iint_{S(m_{ch}, n_{ch}, k_{ch})} \|LK(f_1, f_2, f_1 + f_2 - f)\|^2 \, df_1 df_2$ | (11) |

Which in (11), $S(m_{ch}, n_{ch}, k_{ch})$ is the area confined by three criteria as below:

| | |
|---|---|
| $f_{s,m_{ch}} \le f_1 \le f_{e,m_{ch}}$ | (12) |
| $f_{s,n_{ch}} \le f_2 \le f_{e,n_{ch}}$ | (13) |
| $f'_{s,k_{ch}} \le f_1 + f_2 \le f'_{e,k_{ch}}$ | (14) |

We call the area confined by (12)-(14) an integration island. In figure (1), we can see a typical scheme of the $S(m_{ch}, n_{ch}, k_{ch})$ in the $f_1 - f_2$ plane.

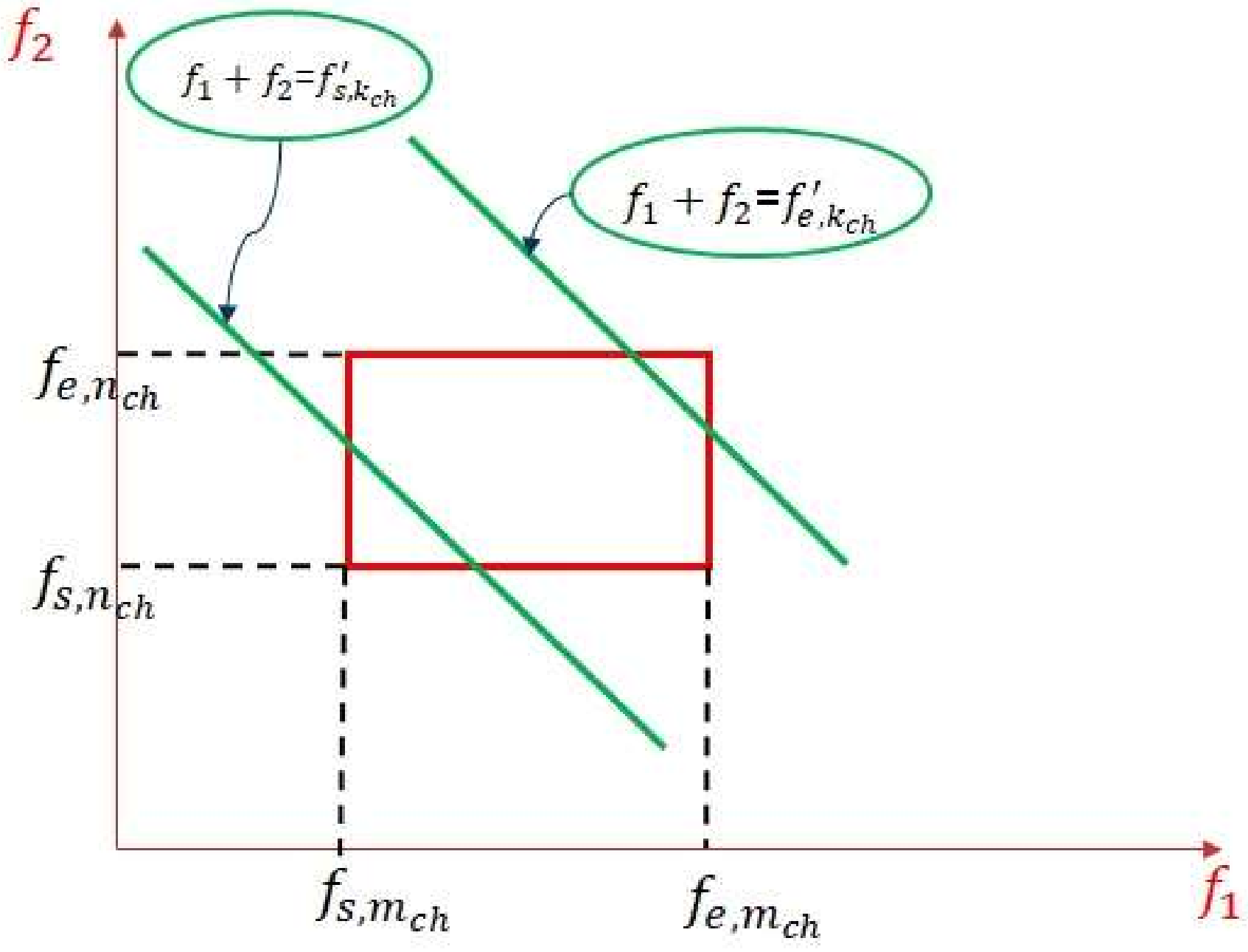

Figure (1): The scheme of the formation of a typical Integration island in f1-f2 plane

As it is evident from figure (1), the integration region is a rectangle that is cut by two parallel lines with slope -1. Based on the values of $f'_{s,k_{ch}}$ and $f'_{e,k_{ch}}$ we may have different shapes for the integration region. Two extreme cases will be:

| | |
|---|---|
| $f'_{s,k_{ch}} > f_{e,m_{ch}} + f_{e,n_{ch}}$ | (15) |
| $f'_{e,k_{ch}} < f_{s,m_{ch}} + f_{s,n_{ch}}$ | (16) |

In fact, when conditions (15) or (16) hold, the integration area vanishes and therefore the value of integral in equation (11) will be zero. The physical meaning of having a null integration region, is that it is impossible for the three channels $m_{ch}, n_{ch}, k_{ch}$ to interact together and form a FWM element in frequency $f$. If (15) and (16) inequalities do not hold, there are several conceivable shapes for integration region that some of them is shown in figure (2).

The diversity of the integration island shapes makes the integral solution in equal (11) very complex particularly when our approach is finding a closed form formula for the 2-D integral. Then, for continue, we need to make an approximation in the shape of the integration islands. Our idea in this case is that we find the geometric center of the integration island and then replace the original shape island with a rectangle whose it's area and geometric center are the same as the original shape.

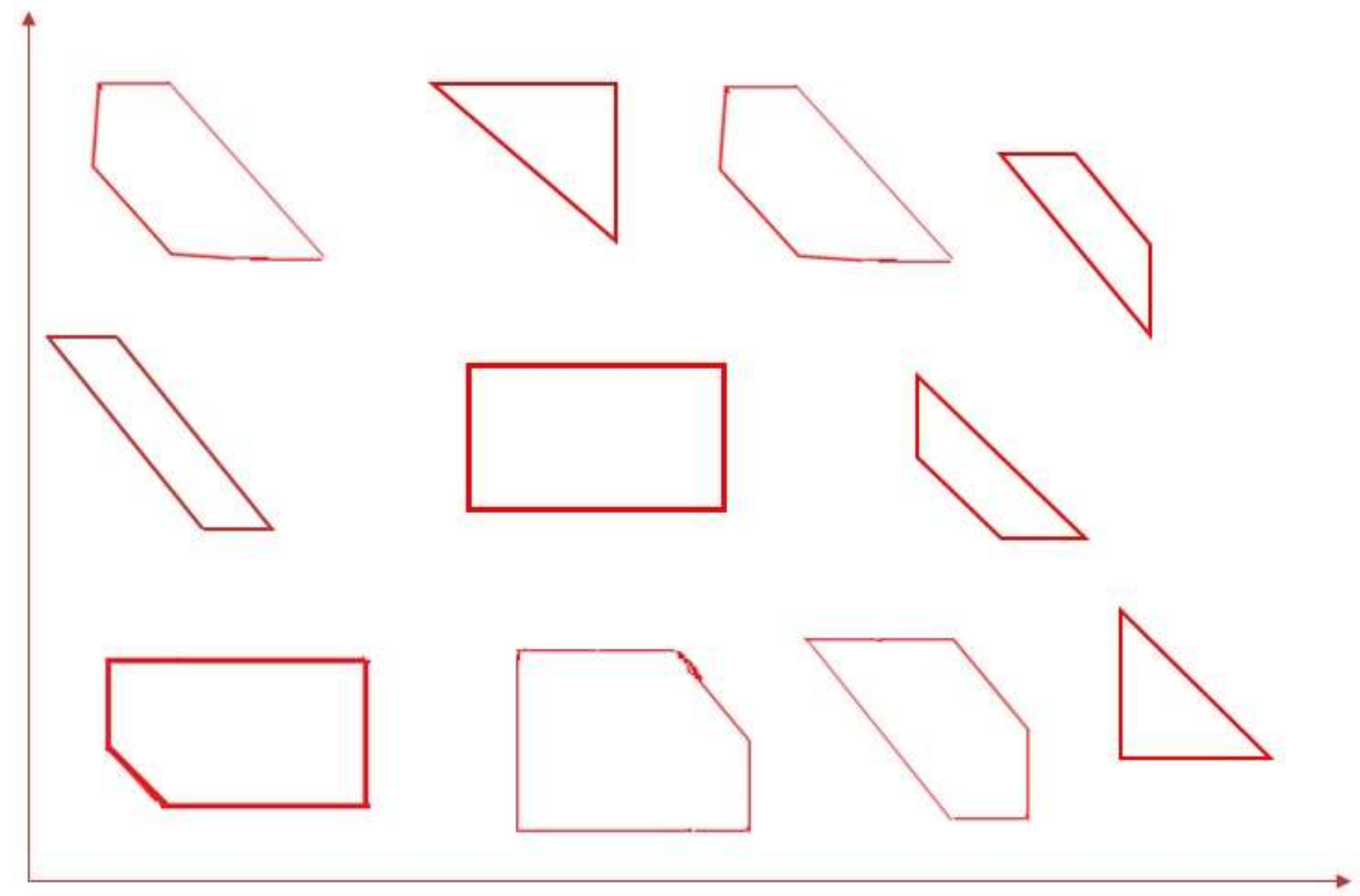

Figure (2): some possible shapes for integration islands

The geometric center points of a right-angled triangle and a rectangle is shown in figure (3).

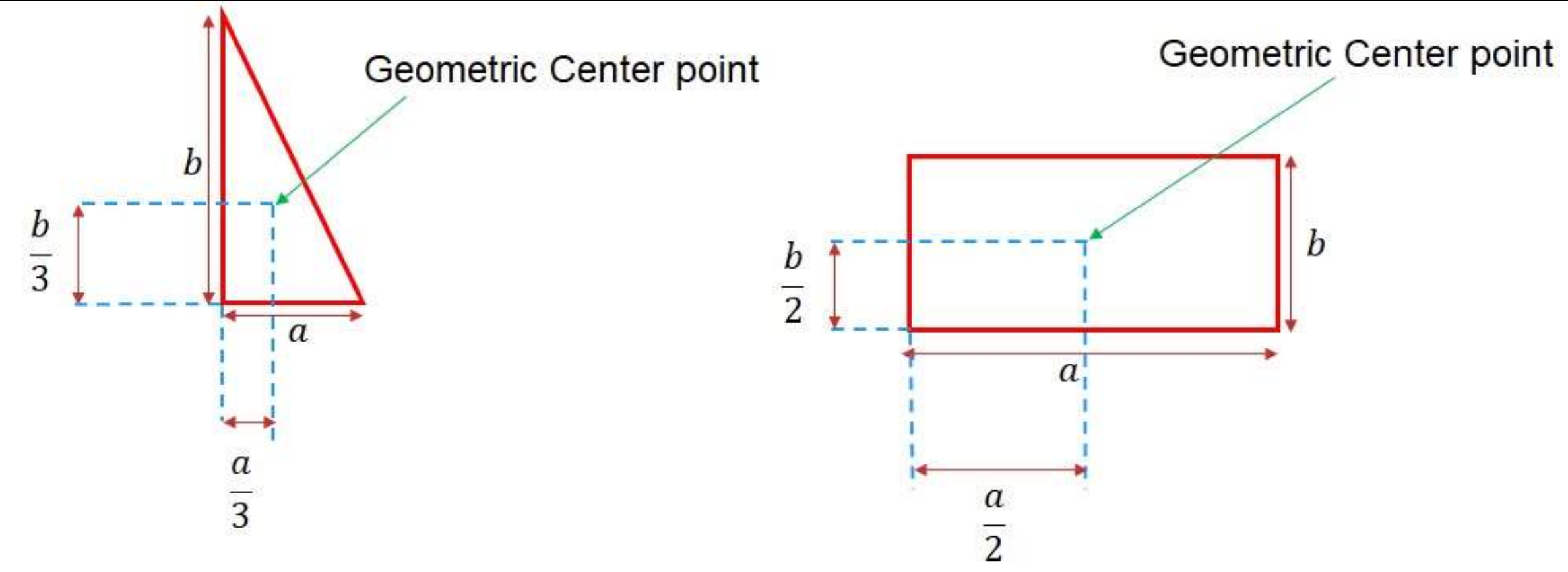

Figure (3): geometric center points of a right-angled triangle and a rectangle

The idea of replacing an arbitrary shape integration island with a concentric equivalent rectangle or square with the same area is depicted in figure (4). As the function under integral is a continuous function, we expect the integration over $S_1$ and $S_2$ are close together and approximately the same. Therefore, the 2-D integral presented in equation (11) can be approximated as:

$$\iint_{S(m_{ch},n_{ch},k_{ch})} |LK(f_1, f_2, f_1 + f_2 - f)|^2 \, df_1 df_2 \cong \iint_{S_{rect}(m_{ch},n_{ch},k_{ch})} |LK(f_1, f_2, f_1 + f_2 - f)|^2 \, df_1 df_2 \quad (17)$$

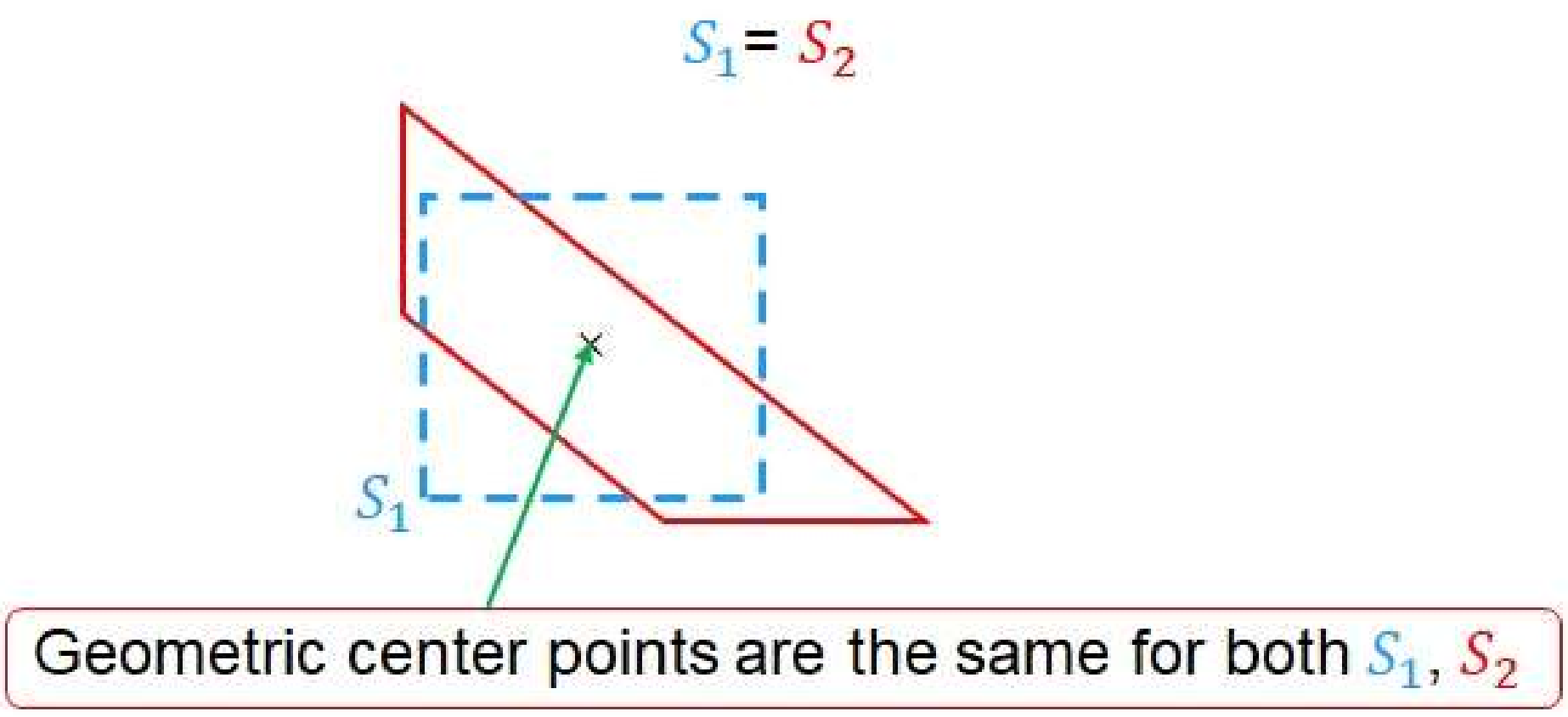

Figure (4): replacing an arbitrary shape integration island with a concentric equivalent rectangle with the same area

Where $S_{rect}(m_{ch}, n_{ch}, k_{ch})$ is the equivalent concentric rectangle while the original integration island shape is $S(m_{ch}, n_{ch}, k_{ch})$.

For finding the area and coordinates of the geometric center of an arbitrary shape island, we investigate 3 specified shapes first and then, we will show that combining the 3 considered shapes we can calculate each arbitrary shape island.

### 3.1- first scenario

The first type of geometry we need, is the one shown in figure (5). We can obviously derive the area and the geometric center of the red color hatched right-angled triangle in figure (5) as:

| | |
|---|---|
| $S_1(\tau) \triangleq \frac{\left(\tau - f_{s,m_{ch}} - f_{s,n_{ch}}\right)^2}{2}$ | (18) |
| $f_1^{(1)}(\tau) \triangleq \frac{2f_{s,m_{ch}}}{3} + \frac{\tau}{3} - \frac{f_{s,n_{ch}}}{3}$ | (19) |
| $f_2^{(1)}(\tau) \triangleq \frac{2f_{s,n_{ch}}}{3} + \frac{\tau}{3} - \frac{f_{s,m_{ch}}}{3}$ | (20) |

### 3.2- second scenario

For the second scenario we may have two situations based on the bandwidth of $m_{ch}$'th and $n_{ch}$'th channels. If the $m_{ch}$'th channel bandwidth is greater or equal to the $n_{ch}$'th channel bandwidth, we have the situation shown in figure (6). On the contrary, if the $n_{ch}$'th channel bandwidth is greater than $m_{ch}$'th channel bandwidth, we have the situation depicted in figure (7).

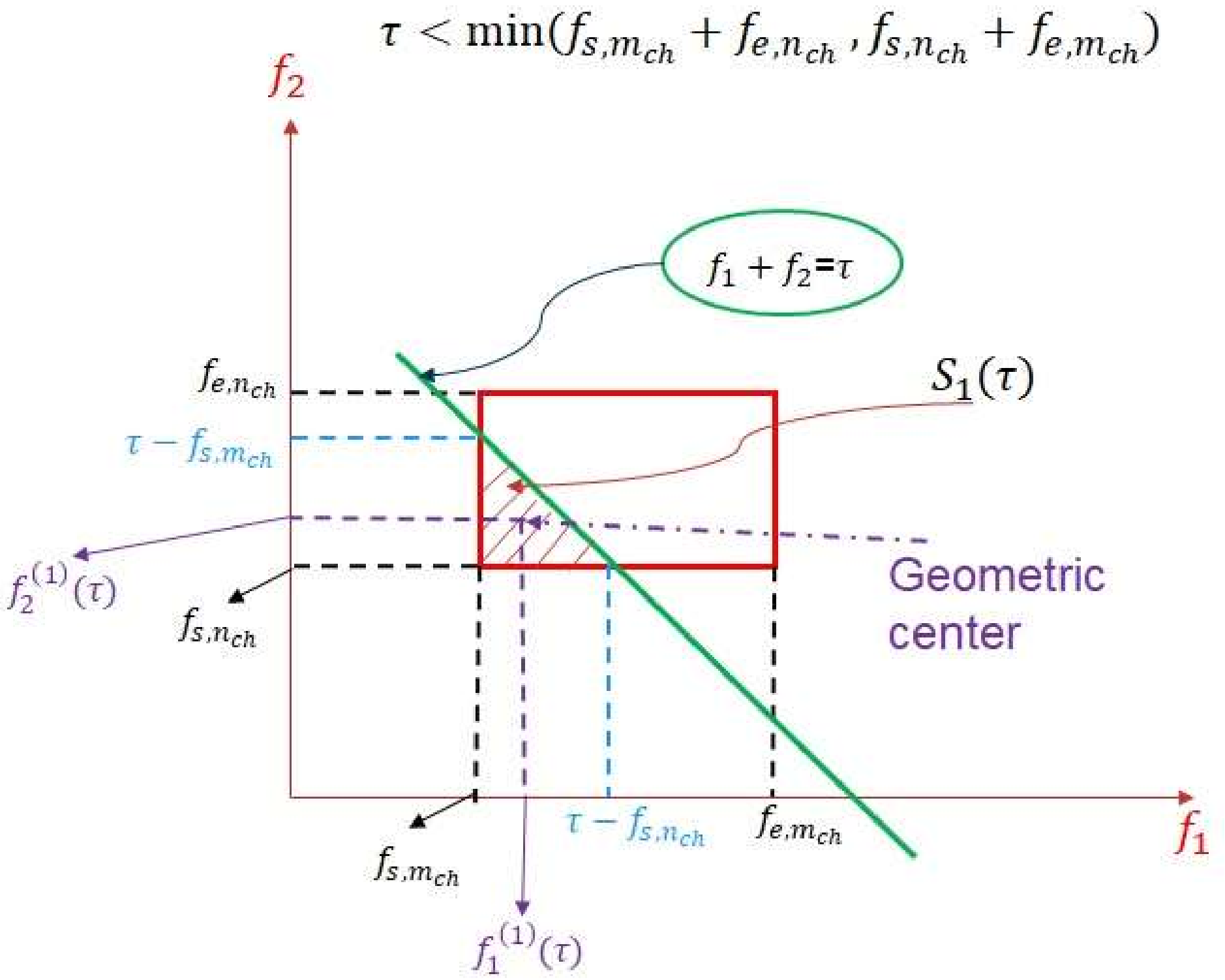

Figure (5): The first basic geometry

Combining the results shown in figures (6) and (7), we have:

| | |
|---|---|
| $BW_{m_{ch}} \triangleq f_{e,m_{ch}} - f_{s,m_{ch}}$ | (21) |
| $BW_{n_{ch}} \triangleq f_{e,n_{ch}} - f_{s,n_{ch}}$ | (22) |
| $u(x) = \begin{cases} 1 & x > 0 \\ \frac{1}{2} & x = 0 \\ 0 & x < 0 \end{cases}$ | (23) |
| $S_2(\tau_1,\tau_2) \triangleq (\tau_2-\tau_1)\times \min(BW_{m_{ch}}, BW_{n_{ch}})$ | (24) |
| $f_1^{(2)}(\tau_1,\tau_2) \triangleq \frac{f_{e,m_{ch}}+f_{s,m_{ch}}}{2}\times u(BW_{n_{ch}}-BW_{m_{ch}}) + \left(\frac{\tau_1+\tau_2}{2}-\frac{f_{e,n_{ch}}+f_{s,n_{ch}}}{2}\right)\times u(BW_{m_{ch}}-BW_{n_{ch}})$ | (25) |
| $f_2^{(2)}(\tau_1,\tau_2) \triangleq \frac{f_{e,n_{ch}}+f_{s,n_{ch}}}{2}\times u(BW_{m_{ch}}-BW_{n_{ch}}) + \left(\frac{\tau_1+\tau_2}{2}-\frac{f_{e,m_{ch}}+f_{s,m_{ch}}}{2}\right)\times u(BW_{n_{ch}}-BW_{m_{ch}})$ | (26) |

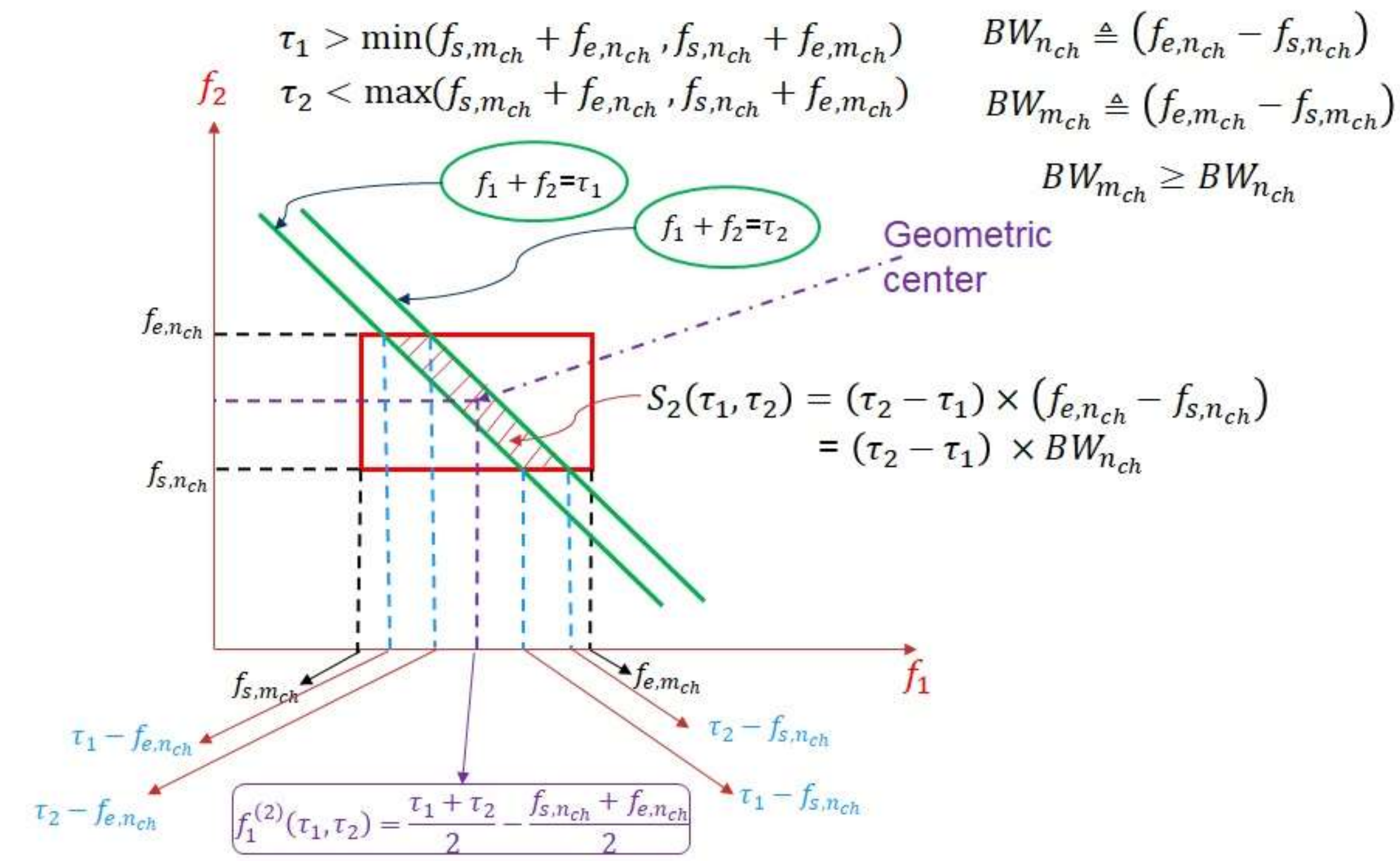

Figure (6): The second basic geometry for $BW_{m_{ch}} > BW_{n_{ch}}$

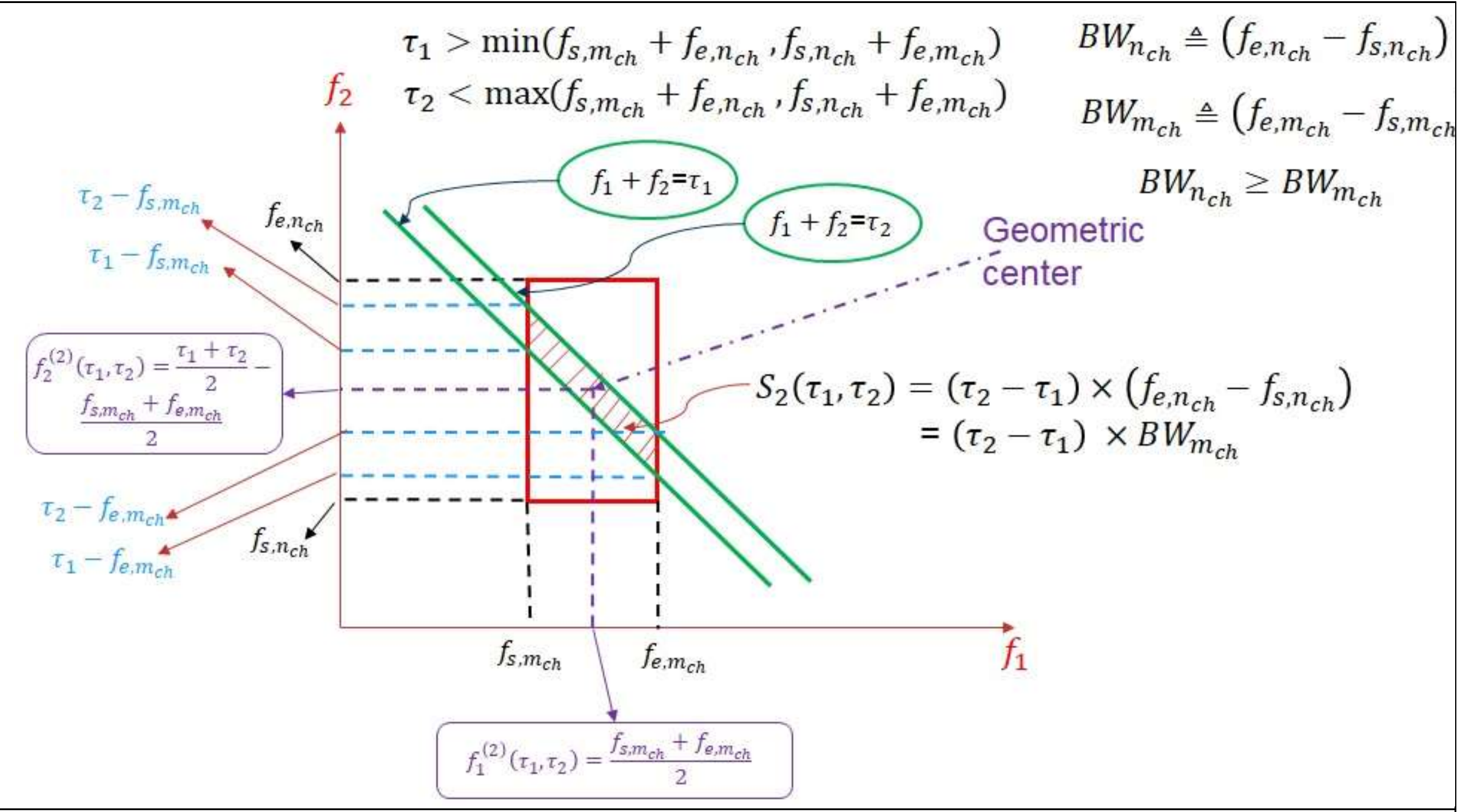

Figure (7): The second basic geometry for $BW_{m_{ch}} < BW_{n_{ch}}$

### 3.3- Third scenario

The 3rd scenario is plotted in figure (8). Based on the coordinates specified in the figure (8), we have the area and geometric center of the hatched right angled triangle in figure (8):

| | |
|---|---|
| $S_3(\tau) \triangleq \frac{(\tau - f_{e,m_{ch}} - f_{e,n_{ch}})^2}{2}$ | (27) |
| $f_1^{(3)}(\tau) \triangleq \frac{2f_{e,m_{ch}}}{3} + \frac{\tau}{3} - \frac{f_{e,n_{ch}}}{3}$ | (28) |
| $f_2^{(3)}(\tau) \triangleq \frac{2f_{e,n_{ch}}}{3} + \frac{\tau}{3} - \frac{f_{e,m_{ch}}}{3}$ | (29) |

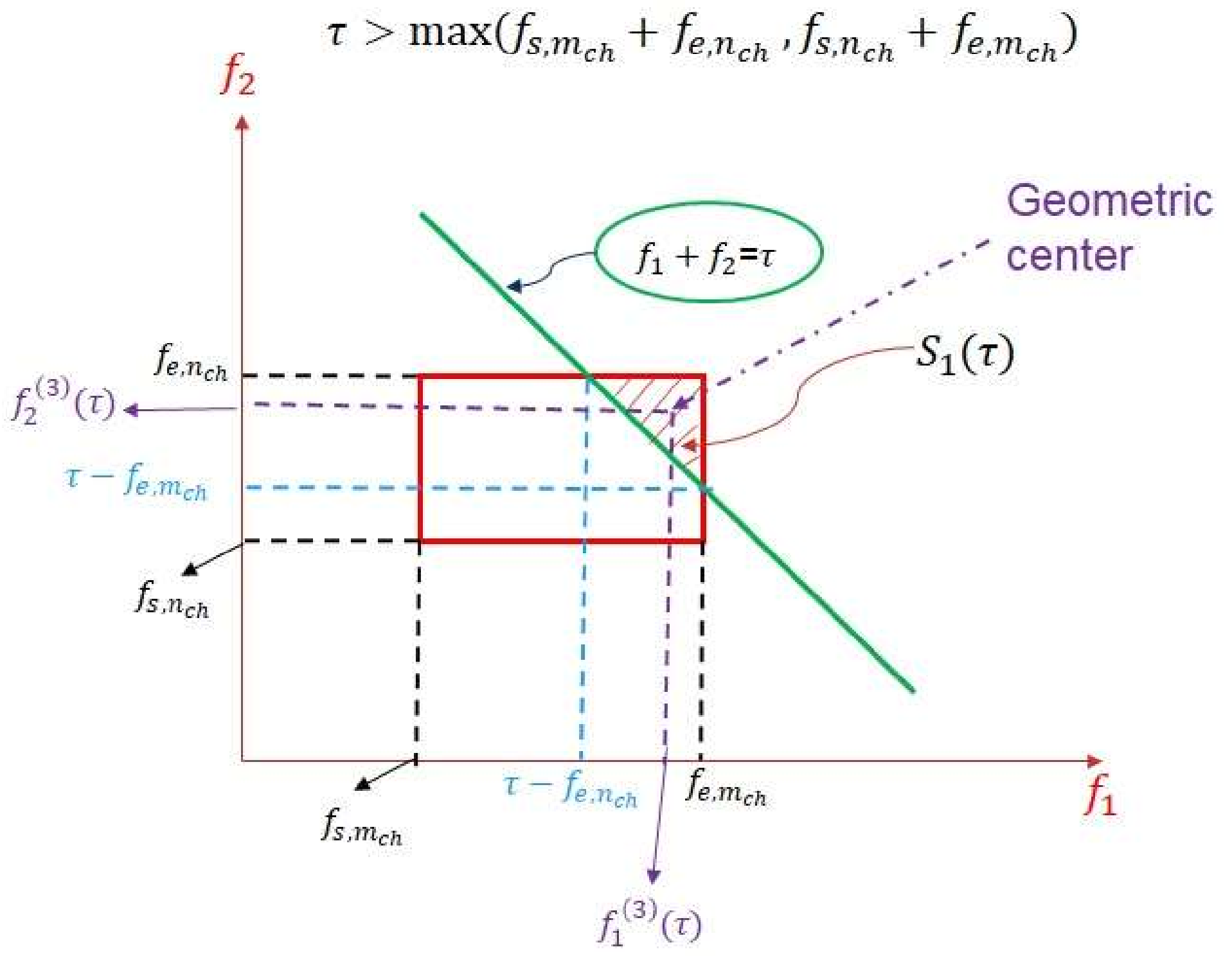

Figure (8): The third basic geometry

There are properties for the geometric center calculation. One important property is that if a big shape formed by the sum of different sub shapes, knowing area and geometric center of each sub shape, we can easily find the geometric center of the big shape as it is shown in figure (9). In general we have:

| | |
|---|---|
| $S = \sum_{k=1}^{n} S_k$ | (30) |
| $f_i^* = \frac{\sum_{k=1}^{n} S_k \times f_i^{k,*}}{\sum_{k=1}^{n} S_k}$ for i = 1.2 | (31) |

On the contrary, if the area and geometric center of the big shape, composed by number of sub shapes, is known and also the area and geometric center of the all the sub shapes except one of them ($k''$th) are known, by using (30)and (31) we can easily deduce for the sub shape with unknown geometric center we have:

| | |
|---|---|
| $S_{k'} = S - \sum_{k=1,k\neq k'}^{n} S_n$ | (32) |
| $f_i^{k',*} = \frac{S \times f_i^* - \sum_{k=1,k\neq k'}^{n} S_k \times f_i^{k,*}}{S - \sum_{k=1,k\neq k'}^{n} S_n}$ for i = 1,2 | (33) |

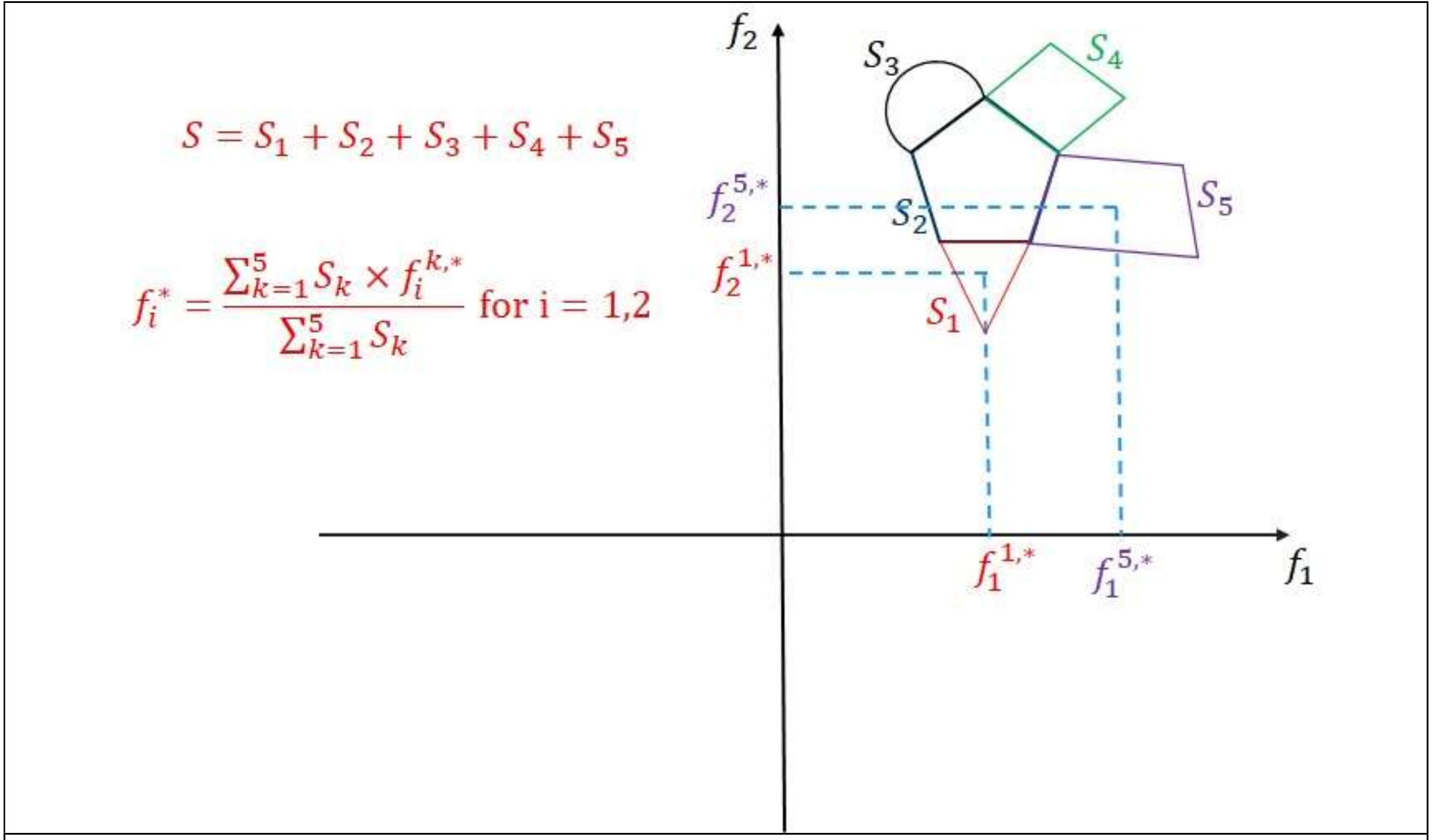

Figure (9): geometric center of a big shape formed by several sub shapes

These properties are useful in derivation of the geometric center of an arbitary shape integration island. In general, if we have three basic typical shape geometric center, we will be able to compute the geometric center of an arbitary island. These general shapes are shown in figure (10).

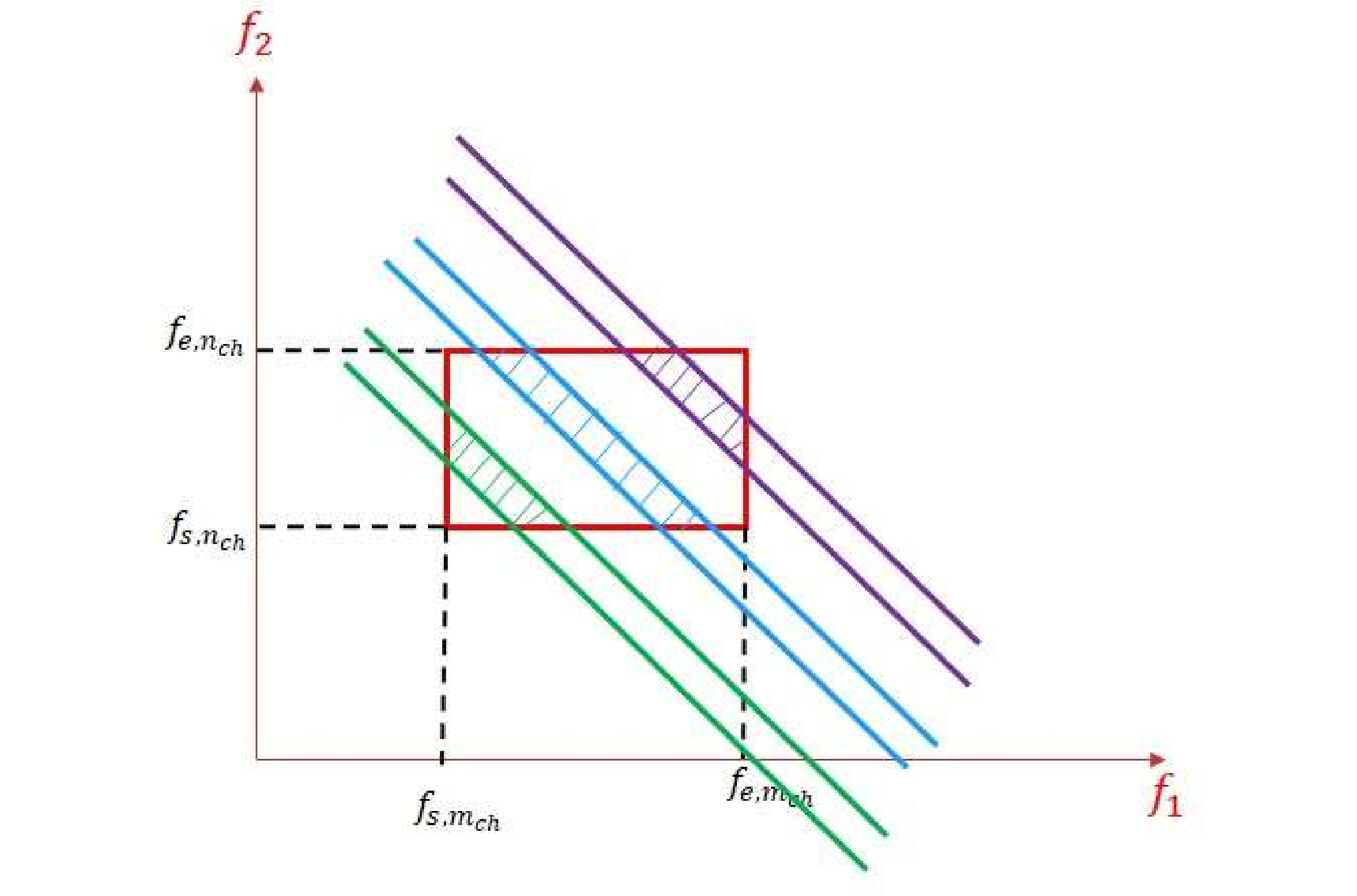

Fig(10): three different general shapesforming an arbitary island

In figure (10), it can be easily seen that the green color hatched area is formed based on the subtraction of a big rigth-angled triangle from a smaller right-angled triangle so the calculation of the area and geometric center of the green color hatched region in figure (10) will be straight forward using equations (18)-(20) and (32)-(33). The blue color hatched region calculations can be done straight forward using (21)-(26). Furthermore, the violet color hatched region area is formed based on the subtraction of two right-angled triangles and using (27)-(29) and (31)-(32) we can find the geometric center and area of the violet hatched region in figure (10). Using formulas (18)-(33) we can summerize the calculation of the area and the geometic center of an arbitary island as below:

| | |
|---|---|
| $F_1 \triangleq f_{s,m_{ch}} + f_{s,n_{ch}}$ | (34) |
| $F_2 \triangleq \min\left(\left(f_{s,m_{ch}} + f_{e,n_{ch}}\right), \left(f_{e,m_{ch}} + f_{s,n_{ch}}\right)\right)$ | (35) |
| $F_3 \triangleq \max\left(\left(f_{s,m_{ch}} + f_{e,n_{ch}}\right), \left(f_{e,m_{ch}} + f_{s,n_{ch}}\right)\right)$ | (36) |
| $F_4 \triangleq f_{e,m_{ch}} + f_{e,n_{ch}}$ | (37) |
| $\tau_1^+ \triangleq \min(f'_{e,k_{ch}}, F_2)$ | (38) |
| $\tau_1^- \triangleq \max(f'_{s,k_{ch}}, F_1)$ | (39) |
| $\tau_1 \triangleq \max(f'_{s,k_{ch}}, F_2)$ | (40) |
| $\tau_2 \triangleq \min(f'_{e,k_{ch}}, F_3)$ | (41) |
| $\tau_3^+ \triangleq \max(f'_{s,k_{ch}}, F_3)$ | (42) |
| $\tau_3^- \triangleq \min(f'_{e,k_{ch}}, F_4)$ | (43) |
| $S_1^+ \triangleq S_1(\tau_1^+) \times U\left(F_2 - f'_{s,k_{ch}}\right) \times U\left(f'_{e,k_{ch}} - F_1\right)$ | (44) |
| $S_1^- \triangleq S_1(\tau_1^-) \times U\left(F_2 - f'_{s,k_{ch}}\right) \times U\left(f'_{e,k_{ch}} - F_1\right)$ | (45) |
| $S_2 \triangleq S_2(\tau_1, \tau_2) \times U\left(F_3 - f'_{s,k_{ch}}\right) \times U\left(f'_{e,k_{ch}} - F_2\right)$ | (46) |
| $S_3^+ \triangleq S_3(\tau_3^+) \times U\left(F_4 - f'_{s,k_{ch}}\right) \times U\left(f'_{e,k_{ch}} - F_3\right)$ | (47) |
| $S_3^- \triangleq S_3(\tau_3^-) \times U\left(F_4 - f'_{s,k_{ch}}\right) \times U\left(f'_{e,k_{ch}} - F_3\right)$ | (48) |
| $S = S_1^+ - S_1^- + S_2 + S_3^+ - S_3^-$ | (49) |
| $L_1 = L_2 = \sqrt{S}$ | (50) |
| $f_1^* = \dfrac{S_1^+ \times f_1^{(1)}(\tau_1^+) - S_1^- \times f_1^{(1)}(\tau_1^-) + S_2 \times f_1^{(2)}(\tau_1, \tau_2) + S_3^+ \times f_1^{(3)}(\tau_3^+) - S_3^- \times f_1^{(3)}(\tau_3^-)}{S_1^+ - S_1^- + S_2 + S_3^+ - S_3^-}$ | (51) |
| $f_2^* = \dfrac{S_1^+ \times f_2^{(1)}(\tau_1^+) - S_1^- \times f_2^{(1)}(\tau_1^-) + S_2 \times f_2^{(2)}(\tau_1, \tau_2) + S_3^+ \times f_2^{(3)}(\tau_3^+) - S_3^- \times f_2^{(3)}(\tau_3^-)}{S_1^+ - S_1^- + S_2 + S_3^+ - S_3^-}$ | (52) |

Using above mentioned equations, we will have the geometric center of an arbitary island and also we have it's area. Note that if the calculated area becomes equal to zero ($S = 0$ eq.49 ), the 2-D integral value for the zero area island will be zero. If $S$ is grater than zero, so the approximated island will be a square with the edge lengths equal to $\sqrt{S}$. Note that, instead of considering a square with the edge lengths equal to $\sqrt{S}$, we can have a rectangle with unequal edge lengths $L_1$ and $L_2$ in general that $S = L_1 \times L_2$. It gives us one more degree of freedom. We leave choosing prper values for $L_1$ and $L_2$ for future research steps.

Therefore, the integral presented in equal (17) can be approximated as:

| | |
|---|---|
| $\iint_{S(m_{ch}, n_{ch}, k_{ch})} \lvert LK(f_1, f_2, f_1 + f_2 - f)\rvert^2 \, df_1 df_2 \cong$ | (53) |

$$\int_{f_2^*-\frac{L_2}{2}}^{f_2^*+\frac{L_2}{2}} \int_{f_1^*-\frac{L_1}{2}}^{f_1^*+\frac{L_1}{2}} |LK(f_1, f_2, f_1+f_2-f)|^2 \, df_1 df_2$$

Where in (53), $f_1^*, f_2^*, L_1, L_2$ are functions of $m_{ch}, n_{ch}, k_{ch}$ due to equations (50)-(52). This functionality is omitted in (53) due to simplicity. At the rest of this paper everywhere we used $f_1^*, f_2^*, L_1, L_2$ they are functions of $m_{ch}, n_{ch}, k_{ch}$ and are obtained using (50)-(52). Finally, having equations (5), (11) and (53) the GN formula can be written as:

$$G_{NLI}(f) \cong \frac{16}{27} \sum_{m_{ch}=1}^{N_c} \sum_{n_{ch}=1}^{N_c} \sum_{k_{ch}=1}^{N_c} G_{m_{ch}} G_{n_{ch}} G_{k_{ch}} \times \int_{f_2^*-\frac{L_2}{2}}^{f_2^*+\frac{L_2}{2}} \int_{f_1^*-\frac{L_1}{2}}^{f_1^*+\frac{L_1}{2}} |LK(f_1, f_2, f_1+f_2-f)|^2 \, df_1 df_2 \tag{54}$$

In the next section, we focus on the integration of the absolute squared link function over an arbitary rectangle region.

## 4. Analytic Integration

A general scheme of a typical span in the optical fiber link is shown in figure (11).

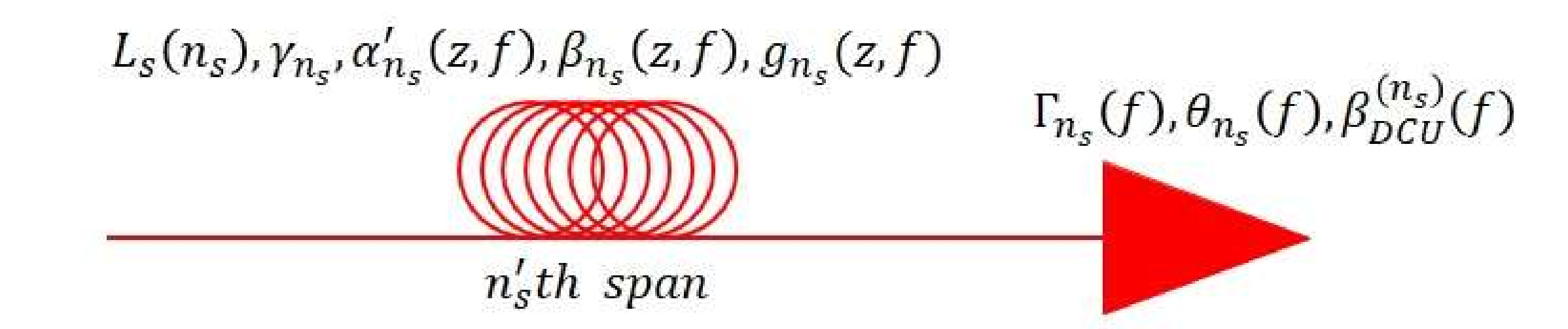

Figure (11): typical scheme of a fiber span

In figure (11), $L_s(n_s)$ is the physical length of the $n'_s th$ fiber span. $\gamma_{n_s}$is the nonlinearity parameter of the $n'_s th$ fiber span which is assumed to be a constant with respect to z (distance) and f (frequency) in one span but it can be different constant values span by span. $\alpha'_{n_s}(z,f)$ and $g_{n_s}(z,f)$ are frequency and distant ($z$) dependent loss and gain respectively through the $n'_s th$ fiber span. The effects of Stimulated Raman Scattering (SRS) can be modeled by $\alpha'_{n_s}(z,f)$ and distributed Raman gain can be modeled through $g_{n_s}(z,f)$.

Also $\beta_{n_s}(z,f)$ in figure (11), is the propagation constant for the $n'_s th$ fiber span which handles all orders of dispersion. In this work we assume that the dispersion is only a function of frequency and not function of the distant so $\beta_{n_s}(z,f)= \beta_{n_s}(f)$ . $\Gamma_{n_s}(f)$ is the power gain of the EDFA at the end of the $n'_s th$ fiber span and $\theta_{n_s}(f)$ is the phase imposed to the electrical field through the EDFA possible linear filtering property. $\beta_{DCU}^{(n_s)}(f)$ is the lumped accumulated dispersion at the end of the $n'_s th$ fiber span. The electrical field input output relation for the amplifier can be written in the frequency domain as:

| | |
|---|---|
| $E_{out}(f) = E_{in}(f)\left(\Gamma_{n_s}(f)\right)^{\frac{1}{2}} e^{j\theta_{n_s}(f)} e^{-j\beta_{DCU}^{(n_s)}(f)}$ | (55) |

The generalized propagation constant for the $n_s'th$ fiber span is defined as [6]:

| | |
|---|---|
| $\kappa_{n_s}(z,f) \triangleq -j\beta_{n_s}(z,f) - \alpha'_{n_s}(z,f) + g_{n_s}(z,f)$ | (56) |

For notation simplicity we define:

| | |
|---|---|
| $\alpha_{n_s}(z,f) \triangleq \alpha'_{n_s}(z,f) - g_{n_s}(z,f)$ | (57) |

Where $\alpha_{n_s}(z,f)$ is the total loss for the $n_s'th$ fiber span. So we have:

| | |
|---|---|
| $\kappa_{n_s}(z,f) = -j\beta_{n_s}(z,f) - \alpha_{n_s}(z,f)$ | (58) |

With the above definitions the link function can be written as [6]:

$$LK(f_1,f_2,f_3) = -j\sum_{n_s=1}^{N_s}\gamma_{n_s}\int_0^{L_s(n_s)} e^{\int_0^{z'}\left[\kappa_{n_s}(z'',\ f_1)+\kappa_{n_s}^*(z'',f_3)+\kappa_{n_s}(z'',f_2)-\kappa_{n_s}(z'',f_1+f_2-f_3)\right]dz''}\,dz'$$
$$\prod_{p=n_s}^{N_s}\Gamma_p^{\frac{1}{2}}(f_1+f_2-f_3)\, e^{j\theta_p(f_1+f_2-f_3)}\, e^{\int_0^{Ls(p)}\kappa_p(z,f_1+f_2-f_3)\,dz}\, e^{-j\beta_{DCU}^{(p)}(f_1+f_2-f_3)}$$
$$\prod_{p=1}^{n_s-1}\Big\{\left[\Gamma_p(f_1)\Gamma_p(f_2)\Gamma_p(f_3)\right]^{\frac{1}{2}} e^{j\left[\theta_p(f_1)+\theta_p(f_2)-\theta_p(f_3)\right]}\, e^{\int_0^{Ls(p)}\left[\kappa_p(z,f_1)+\kappa_p(z,f_2)+\kappa_p^*(z,f_3)\right]dz} \times e^{-j\left[\beta_{DCU}^{(p)}(f_1)+\beta_{DCU}^{(p)}(f_2)-\beta_{DCU}^{(p)}(f_3)\right]}\Big\} \quad (59)$$

We can rewrite equation (59) as:

$$LK(f_1,f_2,f_3) = -j\left(\prod_{p=1}^{N_s} e^{j\theta_p(f_1+f_2-f_3)}\right)\times$$
$$\sum_{n_s=1}^{N_s}\gamma_{n_s}\int_0^{L_s(n_s)} e^{\int_0^{z'}\left[\kappa_{n_s}(z'',\ f_1)+\kappa_{n_s}^*(z'',f_3)+\kappa_{n_s}(z'',f_2)-\kappa_{n_s}(z'',f_1+f_2-f_3)\right]dz''}\,dz'$$
$$\prod_{p=n_s}^{N_s}\left(\Gamma_p(f_1+f_2-f_3)\right)^{\frac{1}{2}} e^{\int_0^{Ls(p)}\kappa_p(z,f_1+f_2-f_3)\,dz}\, e^{-j\beta_{DCU}^{(p)}(f_1+f_2-f_3)}$$
$$\prod_{p=1}^{n_s-1}\Big\{\left[\Gamma_p(f_1)\Gamma_p(f_2)\Gamma_p(f_3)\right]^{\frac{1}{2}} e^{j\left[\theta_p(f_1)+\theta_p(f_2)-\theta_p(f_3)-\theta_p(f_1+f_2-f_3)\right]}$$
$$\times\, e^{\int_0^{Ls(p)}\left[\kappa_p(z,f_1)+\kappa_p(z,f_2)+\kappa_p^*(z,f_3)\right]dz} \times e^{-j\left[\beta_{DCU}^{(p)}(f_1)+\beta_{DCU}^{(p)}(f_2)-\beta_{DCU}^{(p)}(f_3)\right]}\Big\} \quad (60)$$

We assume that the frequency response of the EDFA amplifier is a slowly varying function and remain approximately constant over an integration island. Therefore (60) can be written as:

$$LK(f_1,f_2,f_3) \cong -j\left(\prod_{p=1}^{N_s} e^{j\theta_p(f_1^*+f_2^*-f_3^*)}\right)\times\left(\prod_{p=1}^{N_s} e^{-j\beta_{DCU}^{(p)}(f_1+f_2-f_3)}\right)\times \quad (61)$$

$$\sum_{n_s=1}^{N_s} \gamma_{n_s} \int_0^{L_s(n_s)} e^{\int_0^{z'} [\kappa_{n_s}(z'', f_1)+\kappa^*_{n_s}(z'',f_3)+\kappa_{n_s}(z'',f_2)-\kappa_{n_s}(z'',f_1+f_2-f_3)]\, dz''}\, dz'$$

$$\prod_{p=n_s}^{N_s} \left(\Gamma_p(f_1^*+f_2^*-f_3^*)\right)^{\frac{1}{2}} e^{\int_0^{L_s(p)} \kappa_p(z,f_1+f_2-f_3)\, dz} \times$$

$$\prod_{p=1}^{n_s-1} \Big\{ \left[\Gamma_p(f_1^*)\Gamma_p(f_2^*)\Gamma_p(f_3^*)\right]^{\frac{1}{2}} e^{j[\theta_p(f_1^*)+\theta_p(f_2^*)-\theta_p(f_3^*)-\theta_p(f_1^*+f_2^*-f_3^*)]}$$

$$\times\ e^{\int_0^{L_s(p)} [\kappa_p(z,f_1)+\kappa_p(z,f_2)+\kappa_p^*(z,f_3)]\, dz} \times e^{-j\left[\beta_{DCU}^{(p)}(f_1)+\beta_{DCU}^{(p)}(f_2)-\beta_{DCU}^{(p)}(f_3)-\beta_{DCU}^{(p)}(f_1+f_2-f_3)\right]} \Big\}$$

Where $f_3, f_3^*$ in (61) are:

$$f_3 = f_1 + f_2 - f \qquad f_3^* = f_1^* + f_2^* - f \tag{62}$$

Here, we make an assumption on the total loss function, i.e. $\alpha_{n_s}(z,f)$, which was defined in equation (57). We assume [3]:

$$\alpha_{n_s}(z,f) = \alpha_{0,n_s}(f) + \alpha_{1,n_s}(f) \times \exp(-\sigma_{n_s}(f) \times z) \tag{63}$$

Also for the propagation constant we only consider the Tylor series expansion in frequency domain up to order3.

$$\beta_{n_s}(z,f) = \beta_{n_s}(f) = \beta_{0,n_s} + 2\pi\beta_{1,n_s}(f - f_{n_s}^c) + 4\pi^2 \frac{\beta_{2,n_s}}{2}(f - f_{n_s}^c)^2 + 8\pi^3 \frac{\beta_{3,n_s}}{6}(f - f_{n_s}^c)^3 \tag{64}$$

In equation (64), $f_{n_s}^c$ is the center frequency for Tylor expansion and in general can be different span by span but during each span it must be constant. Also $\beta_{0,n_s}$ , $\beta_{1,n_s}$ , $\beta_{2,n_s}$ , $\beta_{3,n_s}$ are constant values along each span but they can change span by span.

For lumped dispersion element we consider only the second order contribution in Tylor series:

$$\beta_{DCU}^{(p)}(f) = 4\pi^2 \frac{\beta_{DCU,p}}{2}(f - f_{n_s}^c)^2 \tag{65}$$

Where $\beta_{DCU,p}$ is a constant value for each span but it can change from span to span.

Based on equations (58), (63) and (64) we can rewrite the generalized propagation constant as:

$$\kappa_{n_s}(z,f) = -\alpha_{0,n_s}(f) - \alpha_{1,n_s}(f) \times \exp(-\sigma_{n_s}(f) \times z) - j\beta_{0,n_s} - j2\pi\beta_{1,n_s}(f - f_{n_s}^c) - j4\pi^2 \frac{\beta_{2,n_s}}{2}(f - f_{n_s}^c)^2 - j8\pi^3 \frac{\beta_{3,n_s}}{6}(f - f_{n_s}^c)^3 \tag{66}$$

Furthermore, we assume $\alpha_{0,n_s}$ , $\alpha_{1,n_s}$ and $\sigma_{n_s}$ are slowly varying functions of frequency and over an integration island they are approximately constant and their values are approximately the same to the geometric center point of the island. So we have:

| | |
|---|---|
| $\kappa_{n_s}(z,f_i) \cong -\alpha_{0,n_s}(f_i^*) - \alpha_{1,n_s}(f_i^*) \times \exp\left(-\sigma_{n_s}(f_i^*) \times z\right) - j\beta_{0,n_s} - j2\pi\beta_{1,n_s}\left(f_i - f_{n_s}^c\right) - j4\pi^2 \frac{\beta_{2,n_s}}{2}\left(f_i - f_{n_s}^c\right)^2 - j8\pi^3 \frac{\beta_{3,n_s}}{6}\left(f_i - f_{n_s}^c\right)^3 ; i = 1,2,3,4$ | (67) |

Where:

| | |
|---|---|
| $f_4 \triangleq f_1 + f_2 - f_3 = f$ <br> $f_4^* \triangleq f_1^* + f_2^* - f_3^* = f$ | (68) |

To continue, we first define:

| | |
|---|---|
| $\overline{\alpha_{0,n_s}} \triangleq \frac{\alpha_{0,n_s}(f_1^*) + \alpha_{0,n_s}(f_2^*) + \alpha_{0,n_s}(f_3^*) - \alpha_{0,n_s}(f)}{2}$ | (69) |

And also we assume that:

| | |
|---|---|
| $\overline{\alpha_{1,n_s}} exp\left(-\overline{\sigma_{n_s}} \times z\right) \cong \frac{1}{2}\left\{\alpha_{1,n_s}(f_1^*) \times \exp\left(-\sigma_{n_s}(f_1^*) \times z\right) + \alpha_{1,n_s}(f_2^*) \times \exp\left(-\sigma_{n_s}(f_2^*) \times z\right) - \alpha_{1,n_s}(f) \times \exp\left(-\sigma_{n_s}(f) \times z\right) + \alpha_{1,n_s}(f_1^* + f_2^* - f) \times \exp\left(-\sigma_{n_s}(f_1^* + f_2^* - f) \times z\right)\right\} ; for\ 0 \leq z \leq L_s(n_s)$ | (70) |

Our basic assumption in (70) is that we can find suitable values for $\overline{\alpha_{1,n_s}}$ , $\overline{\sigma_{n_s}}$ such that equation (17) holds with a good approximation for all values of $0 \leq z \leq L_s(n_s)$ where $L_s(n_s)$ is the length of the $n_s'th$ fiber span. By using (67), (69) and (70):

| | |
|---|---|
| $\kappa_{n_s}(z'',f_1) + \kappa_{n_s}^*(z'',f_3) + \kappa_{n_s}(z'',f_2) - \kappa_{n_s}(z'',f_1 + f_2 - f_3) = -2\overline{\alpha_{0,n_s}} - 2\overline{\alpha_{1,n_s}} exp\left(-\overline{\sigma_{n_s}} \times z''\right) + j4\pi^2(f_1 - f)(f_2 - f) \times \left(\beta_{2,n_s} + \pi\beta_{3,n_s}\left(f_1 + f_2 - 2f_{n_s}^c\right)\right)$ | (71) |

Therefore we can easily calculate:

| | |
|---|---|
| $\int_0^{z'} \left[\kappa_{n_s}(z'', f_1) + \kappa_{n_s}^*(z'',f_3) + \kappa_{n_s}(z'',f_2) - \kappa_{n_s}(z'',f_1 + f_2 - f_3)\right] dz'' = -2\overline{\alpha_{0,n_s}} z' + \frac{2\overline{\alpha_{1,n_s}}}{\overline{\sigma_{n_s}}} exp\left(-\overline{\sigma_{n_s}} \times z'\right) - \frac{2\overline{\alpha_{1,n_s}}}{\overline{\sigma_{n_s}}} + j4\pi^2(f_1 - f)(f_2 - f)z' \times \left(\beta_{2,n_s} + \pi\beta_{3,n_s}\left(f_1 + f_2 - 2f_{n_s}^c\right)\right)$ | (72) |

Let's assume $\overline{\alpha_{1,n_s}} = 0$, so we will have:

| | |
|---|---|
| $\int_0^{L_s(n_s)} e^{\int_0^{z'}\left[\kappa_{n_s}(z'', f_1) + \kappa_{n_s}^*(z'',f_3) + \kappa_{n_s}(z'',f_2) - \kappa_{n_s}(z'',f_1+f_2-f_3)\right] dz''} dz' =$ | (73) |

$$\int_0^{L_s(n_s)} e^{z'\left(-2\overline{\alpha_{0,n_s}}+j4\pi^2(f_1-f)(f_2-f)\left(\beta_{2,n_s}+\pi\beta_{3,n_s}\left(f_1+f_2-2f_{n_s}^c\right)\right)\right)}\, dz' = \frac{e^{L_s(n_s)\left(-2\overline{\alpha_{0,n_s}}+j4\pi^2(f_1-f)(f_2-f)\left(\beta_{2,n_s}+\pi\beta_{3,n_s}\left(f_1+f_2-2f_{n_s}^c\right)\right)\right)}-1}{-2\overline{\alpha_{0,n_s}}+j4\pi^2(f_1-f)(f_2-f)\left(\beta_{2,n_s}+\pi\beta_{3,n_s}\left(f_1+f_2-2f_{n_s}^c\right)\right)},\quad if\ \overline{\alpha_{1,n_s}}=0$$

Another reasonable assumption is that $e^{-2\overline{\alpha_{0,n_s}}\times L_s(n_s)} \ll 1$ which notifies each span loss is high enough. Therefore we will have:

$$\int_0^{L_s(n_s)} e^{\int_0^{z'}\left[\kappa_{n_s}(z'',\ f_1)+\kappa_{n_s}^*(z'',f_3)+\kappa_{n_s}(z'',f_2)-\kappa_{n_s}(z'',f_1+f_2-f_3)\right]dz''}\, dz' \cong \frac{1}{2\overline{\alpha_{0,n_s}}-j4\pi^2(f_1-f)(f_2-f)\left(\beta_{2,n_s}+\pi\beta_{3,n_s}\left(f_1+f_2-2f_{n_s}^c\right)\right)},\quad if\ \overline{\alpha_{1,n_s}}=0 \qquad (74)$$

When $\overline{\alpha_{1,n_s}} \neq 0$, the exact analytical solution of the integral in equation (73) is very complicated but under the condition that $\overline{\alpha_{1,n_s}} \ll \overline{\alpha_{0,n_s}}$ we will have the approximate integral analytic solution as [3]:

$$\int_0^{L_s(n_s)} e^{\int_0^{z'}\left[\kappa_{n_s}(z'',\ f_1)+\kappa_{n_s}^*(z'',f_3)+\kappa_{n_s}(z'',f_2)-\kappa_{n_s}(z'',f_1+f_2-f_3)\right]dz''}\, dz' \cong \frac{1}{2\overline{\alpha_{0,n_s}}-j4\pi^2(f_1-f)(f_2-f)\left(\beta_{2,n_s}+\pi\beta_{3,n_s}\left(f_1+f_2-2f_{n_s}^c\right)\right)} - \left\{\frac{2\overline{\alpha_{1,n_s}}}{\left[2\overline{\alpha_{0,n_s}}-j4\pi^2(f_1-f)(f_2-f)\left(\beta_{2,n_s}+\pi\beta_{3,n_s}\left(f_1+f_2-2f_{n_s}^c\right)\right)\right]}\times \frac{1}{\left[2\overline{\alpha_{0,n_s}}+\overline{\sigma_{n_s}}-j4\pi^2(f_1-f)(f_2-f)\left(\beta_{2,n_s}+\pi\beta_{3,n_s}\left(f_1+f_2-2f_{n_s}^c\right)\right)\right]}\right\},\ for\ \overline{\alpha_{1,n_s}} \ll \overline{\alpha_{0,n_s}} \qquad (74)$$

Another integral in equation (61) which must be solved analytically is:

$$\int_0^{L_s(p)} \kappa_p(z\,, f_1+f_2-f_3)\ dz = \int_0^{L_s(p)} \left(-\alpha_{0,n_s}(f)-\alpha_{1,n_s}(f)\times\exp\left(-\sigma_{n_s}(f)\times z\right)-j\beta_{n_s}(f)\right)dz = \left(-\alpha_{0,n_s}(f)-j\beta_{n_s}(f)\right)L_s(p)+\frac{\alpha_{1,n_s}(f)}{\sigma_{n_s}(f)}\exp\left(-\sigma_{n_s}(f)\times L_s(p)\right)-\frac{\alpha_{1,n_s}(f)}{\sigma_{n_s}(f)} \qquad (75)$$

So we have:

| | |
|---|---|
| $e^{\int_0^{L_s(p)} \kappa_p(z, f_1+f_2-f_3)\, dz} = e^{-\left(\alpha_{0,n_s}(f) L_s(p) + \frac{\alpha_{1,n_s}(f)}{\sigma_{n_s}(f)} - \frac{\alpha_{1,n_s}(f)}{\sigma_{n_s}(f)} \exp(-\sigma_{n_s}(f) \times L_s(p))\right)} \times e^{-j\beta_{n_s}(f) L_s(p)}$ | (76) |

The last integral in equation (61) which must be solved analytically using (58) and (63) is:

| | |
|---|---|
| $\int_0^{L_s(p)} \left[\kappa_p(z, f_1) + \kappa_p(z, f_2) + \kappa_p^*(z, f_3)\right] dz = -\left(\alpha_{0,n_s}(f_1^*) + \alpha_{0,n_s}(f_2^*) + \alpha_{0,n_s}(f_3^*)\right) L_s(p) - j\left(\beta_{n_s}(f_1) + \beta_{n_s}(f_2) - \beta_{n_s}(f_3)\right) L_s(p) + \sum_{i=1}^{3} \left\{ \frac{\alpha_{1,n_s}(f_i^*)}{\sigma_{n_s}(f_i^*)} \exp\left(-\sigma_{n_s}(f_i^*) \times L_s(p)\right) - \frac{\alpha_{1,n_s}(f_i^*)}{\sigma_{n_s}(f_i^*)} \right\}$ | (77) |

Therefore we have:

| | |
|---|---|
| $e^{\int_0^{L_s(p)} [\kappa_p(z, f_1) + \kappa_p(z, f_2) + \kappa_p^*(z, f_3)]\, dz} = e^{-L_s(p) \sum_{i=1}^{3} \alpha_{0,n_s}(f_i^*)} \times e^{j\left(\beta_{n_s}(f_1) + \beta_{n_s}(f_2) - \beta_{n_s}(f_3)\right) L_s(p)} \times e^{\sum_{i=1}^{3} \left\{ \frac{\alpha_{1,n_s}(f_i^*)}{\sigma_{n_s}(f_i^*)} \exp\left(-\sigma_{n_s}(f_i^*) \times L_s(p)\right) - \frac{\alpha_{1,n_s}(f_i^*)}{\sigma_{n_s}(f_i^*)} \right\}}$ | (78) |

Using (74), (76) and (78), equation (61) can be rewritten as:

| | |
|---|---|
| $LK(f_1, f_2, f_3) \cong -j \left( \prod_{p=1}^{N_s} e^{j\theta_p(f_1^* + f_2^* - f_3^*)} \right) \times \left( \prod_{p=1}^{N_s} e^{-j\beta_{DCU}^{(p)}(f_1+f_2-f_3)} \right) \times \sum_{n_s=1}^{N_s} \gamma_{n_s} \left\{ \frac{1}{2\overline{\alpha_{0,n_s}} - j4\pi^2 (f_1 - f)(f_2 - f)\left(\beta_{2,n_s} + \pi\beta_{3,n_s}(f_1 + f_2 - 2f_{n_s}^c)\right)} - \frac{2\overline{\alpha_{1,n_s}}}{\left[2\overline{\alpha_{0,n_s}} - j4\pi^2 (f_1 - f)(f_2 - f)\left(\beta_{2,n_s} + \pi\beta_{3,n_s}(f_1 + f_2 - 2f_{n_s}^c)\right)\right]} \times \frac{1}{\left[2\overline{\alpha_{0,n_s}} + \overline{\sigma_{n_s}} - j4\pi^2 (f_1 - f)(f_2 - f)\left(\beta_{2,n_s} + \pi\beta_{3,n_s}(f_1 + f_2 - 2f_{n_s}^c)\right)\right]} \right\} \prod_{p=n_s}^{N_s} \left(\Gamma_p(f)\right)^{\frac{1}{2}} \times$ | (79) |

$$\left\{ e^{-(\alpha_{0,n_s}(f)L_s(p)+\frac{\alpha_{1,n_s}(f)}{\sigma_{n_s}(f)}-\frac{\alpha_{1,n_s}(f)}{\sigma_{n_s}(f)}\exp(-\sigma_{n_s}(f)\times L_s(p)))} \times e^{-j\beta_{n_s}(f)L_s(p)} \right\} \times$$

$$\prod_{p=1}^{n_s-1} \left\{ \left[\Gamma_p(f_1^*)\Gamma_p(f_2^*)\Gamma_p(f_3^*)\right]^{\frac{1}{2}} e^{j[\theta_p(f_1^*)+\theta_p(f_2^*)-\theta_p(f_3^*)-\theta_p(f_1^*+f_2^*-f_3^*)]} \right.$$

$$e^{-L_s(p)\sum_{i=1}^{3}\alpha_{0,n_s}(f_i^*)} \times e^{-j\left(\beta_{n_s}(f_1)+\beta_{n_s}(f_2)-\beta_{n_s}(f_3)\right)L_s(p)} \times$$

$$e^{\sum_{i=1}^{3}\left\{\frac{\alpha_{1,n_s}(f_i^*)}{\sigma_{n_s}(f_i^*)}\exp\left(-\sigma_{n_s}(f_i^*)\times L_s(p)\right)-\frac{\alpha_{1,n_s}(f_i^*)}{\sigma_{n_s}(f_i^*)}\right\}}$$

$$\left. \times e^{-j\left[\beta_{DCU}^{(p)}(f_1)+\beta_{DCU}^{(p)}(f_2)-\beta_{DCU}^{(p)}(f_3)-\beta_{DCU}^{(p)}(f_1+f_2-f_3)\right]} \right\}$$

For notation simplicity we define:

$$g_0(n_s) \triangleq \prod_{p=n_s}^{N_s}\left(\Gamma_p(f)\right)^{\frac{1}{2}} \times \left\{ e^{-\left(\alpha_{0,n_s}(f)L_s(p)+\frac{\alpha_{1,n_s}(f)}{\sigma_{n_s}(f)}-\frac{\alpha_{1,n_s}(f)}{\sigma_{n_s}(f)}\exp\left(-\sigma_{n_s}(f)\times L_s(p)\right)\right)} \right\} \times \prod_{p=1}^{n_s-1}\left\{ \left[\Gamma_p(f_1^*)\Gamma_p(f_2^*)\Gamma_p(f_3^*)\right]^{\frac{1}{2}} \times e^{-L_s(p)\sum_{i=1}^{3}\alpha_{0,n_s}(f_i^*)} \times e^{\sum_{i=1}^{3}\left\{\frac{\alpha_{1,n_s}(f_i^*)}{\sigma_{n_s}(f_i^*)}\exp\left(-\sigma_{n_s}(f_i^*)\times L_s(p)\right)-\frac{\alpha_{1,n_s}(f_i^*)}{\sigma_{n_s}(f_i^*)}\right\}} \right\} \quad (80)$$

Note that $g_0(n_s)$ does not have a functionality of $f_1$or $f_2$. Therefore equation (79) can be written as:

$$LK(f_1,f_2,f_3) \cong -j\left(\prod_{p=1}^{N_s} e^{j\theta_p(f_1^*+f_2^*-f_3^*)}\right) \times \left(\prod_{p=1}^{N_s} e^{-j\beta_{DCU}^{(p)}(f_1+f_2-f_3)}\right) \times \left(\prod_{p=1}^{N_s} e^{-j\beta_{n_s}(f)L_s(p)}\right)$$

$$\sum_{n_s=1}^{N_s}\gamma_{n_s}\times g_0(n_s)\left\{\frac{1}{2\overline{\alpha_{0,n_s}}-j4\pi^2(f_1-f)(f_2-f)\left(\beta_{2,n_s}+\pi\beta_{3,n_s}\left(f_1+f_2-2f_{n_s}^c\right)\right)} - \frac{2\overline{\alpha_{1,n_s}}}{\left[2\overline{\alpha_{0,n_s}}-j4\pi^2(f_1-f)(f_2-f)\left(\beta_{2,n_s}+\pi\beta_{3,n_s}\left(f_1+f_2-2f_{n_s}^c\right)\right)\right]} \times \frac{1}{\left[2\overline{\alpha_{0,n_s}}+\overline{\sigma_{n_s}}-j4\pi^2(f_1-f)(f_2-f)\left(\beta_{2,n_s}+\pi\beta_{3,n_s}\left(f_1+f_2-2f_{n_s}^c\right)\right)\right]}\right\}$$

$$\prod_{p=1}^{n_s-1}\left\{ e^{j[\theta_p(f_1^*)+\theta_p(f_2^*)-\theta_p(f_3^*)-\theta_p(f_1^*+f_2^*-f_3^*)]} \times e^{-j\left(\beta_{n_s}(f_1)+\beta_{n_s}(f_2)-\beta_{n_s}(f_3)-\beta_{n_s}(f)\right)L_s(p)} \times e^{-j\left[\beta_{DCU}^{(p)}(f_1)+\beta_{DCU}^{(p)}(f_2)-\beta_{DCU}^{(p)}(f_3)-\beta_{DCU}^{(p)}(f_1+f_2-f_3)\right]} \right\} \quad (81)$$

The link function in equation (81) can be further simplified as:

$$LK(f_1,f_2,f_3) \cong -je^{j\sum_{p=1}^{N_s}\left(\theta_p(f_1^*+f_2^*-f_3^*)-\beta_{DCU}^{(p)}(f_1+f_2-f_3)-\beta_{n_s}(f)L_s(p)\right)} \times \sum_{n_s=1}^{N_s}\gamma_{n_s}\times g_0(n_s)\left\{\frac{1}{2\overline{\alpha_{0,n_s}}-j4\pi^2(f_1-f)(f_2-f)\left(\beta_{2,n_s}+\pi\beta_{3,n_s}(f_1+f_2-2f_{n_s}^c)\right)} - \frac{2\overline{\alpha_{1,n_s}}}{\left[2\overline{\alpha_{0,n_s}}-j4\pi^2(f_1-f)(f_2-f)\left(\beta_{2,n_s}+\pi\beta_{3,n_s}(f_1+f_2-2f_{n_s}^c)\right)\right]}\times \frac{1}{\left[2\overline{\alpha_{0,n_s}}+\overline{\sigma_{n_s}}-j4\pi^2(f_1-f)(f_2-f)\left(\beta_{2,n_s}+\pi\beta_{3,n_s}(f_1+f_2-2f_{n_s}^c)\right)\right]}\right\} e^{j\sum_{p=1}^{n_s-1}\left(\theta_p(f_1^*)+\theta_p(f_2^*)-\theta_p(f_3^*)-\theta_p(f_1^*+f_2^*-f_3^*)\right)}\times e^{-j\sum_{p=1}^{n_s-1}\left(\beta_{n_s}(f_1)+\beta_{n_s}(f_2)-\beta_{n_s}(f_3)-\beta_{n_s}(f)\right)L_s(p)}\times e^{-j\sum_{p=1}^{n_s-1}\left(\beta_{DCU}^{(p)}(f_1)+\beta_{DCU}^{(p)}(f_2)-\beta_{DCU}^{(p)}(f_3)-\beta_{DCU}^{(p)}(f_1+f_2-f_3)\right)} \quad (82)$$

Also for more simplicity we define:

$$\theta_0^{n_s} \triangleq \sum_{p=1}^{n_s-1}\left(\theta_p(f_1^*)+\theta_p(f_2^*)-\theta_p(f_3^*)-\theta_p(f_1^*+f_2^*-f_3^*)\right) \quad (83)$$

Using (64) we have:

$$-j\times\left(\beta_{n_s}(f_1)+\beta_{n_s}(f_2)-\beta_{n_s}(f_3)-\beta_{n_s}(f)\right) = j\times 4\pi^2(f_1-f)(f_2-f)\times \left(\beta_{2,n_s}+\pi\beta_{3,n_s}(f_1+f_2-2f_{n_s}^c)\right) \quad (84)$$

Also using (65), we can write:

$$-j\left(\beta_{DCU}^{(p)}(f_1)+\beta_{DCU}^{(p)}(f_2)-\beta_{DCU}^{(p)}(f_3)-\beta_{DCU}^{(p)}(f_1+f_2-f_3)\right) = = j\times 4\pi^2(f_1-f)(f_2-f)\times\beta_{DCU,p} \quad (84)$$

Having (83) and (84), equation (82) can be written as:

$$LK(f_1,f_2,f_3) \cong -je^{j\sum_{p=1}^{N_s}\left(\theta_p(f_1^*+f_2^*-f_3^*)-\beta_{DCU}^{(p)}(f_1+f_2-f_3)-\beta_{n_s}(f)L_s(p)\right)} \times \sum_{n_s=1}^{N_s}\gamma_{n_s}\times g_0(n_s)\times e^{j\theta_0^{n_s}}\times\xi(n_s,f_1,f_2,f)\times e^{j4\pi^2(f_1-f)(f_2-f)\times\beta_{acc}(n_s,f_1,f_2)} \quad (85)$$

Where $\xi(n_s,f_1,f_2,f)$ in (85) is:

$$\xi(n_s,f_1,f_2,f) \triangleq \left\{\frac{1}{2\overline{\alpha_{0,n_s}}-j4\pi^2(f_1-f)(f_2-f)\left(\beta_{2,n_s}+\pi\beta_{3,n_s}(f_1+f_2-2f_{n_s}^c)\right)} - \frac{2\overline{\alpha_{1,n_s}}}{\left[2\overline{\alpha_{0,n_s}}-j4\pi^2(f_1-f)(f_2-f)\left(\beta_{2,n_s}+\pi\beta_{3,n_s}(f_1+f_2-2f_{n_s}^c)\right)\right]}\times \right. \quad (86)$$

$$\left.\frac{1}{\left[2\overline{\alpha_{0,n_s}}+\overline{\sigma_{n_s}}-j4\pi^2(f_1-f)(f_2-f)\left(\beta_{2,n_s}+\pi\beta_{3,n_s}\left(f_1+f_2-2f_{n_s}^c\right)\right)\right]}\right\}$$

Also $\beta_{acc}(n_s,f_1,f_2)$ in (85) is:

$$\beta_{acc}(n_s,f_1,f_2) \triangleq \sum_{p=1}^{n_s-1}\left[\left(\beta_{2,p}+\pi\beta_{3,p}\left(f_1+f_2-2f_p^c\right)\right)\times L_s(p)+\beta_{DCU,p}\right]$$
$$= \sum_{p=1}^{n_s-1}\left(\beta_{2,p}\times L_s(p)+\beta_{DCU,p}\right)+\pi(f_1+f_2)\sum_{p=1}^{n_s-1}\left(\beta_{3,p}\times L_s(p)\right)$$
$$-2\pi\sum_{p=1}^{n_s-1}\left(\beta_{3,p}\times f_p^c\times L_s(p)\right) \quad (87)$$

Having equation (85), the absolute square of the link function will be:

$$|LK(f_1,f_2,f_3)|^2 \cong \sum_{n_s=1}^{N_s}\sum_{n_s'=1}^{N_s}\gamma_{n_s}\gamma_{n_s'}\times g_0(n_s')g_0(n_s)\times\ e^{j\Delta\theta(n_s,n_s')}\times\xi(n_s,f_1,f_2,f)\times$$
$$\xi^*(n_s',f_1,f_2,f)\times\ e^{j4\pi^2(f_1-f)(f_2-f)\times\Delta\beta_{acc}(n_s,n_s',f_1,f_2)} \quad (88)$$

Where in (88), $\Delta\theta(n_s,n_s')$ and $\Delta\beta_{acc}(n_s,n_s',f_1,f_2)$ are:

$$\Delta\theta(n_s,n_s') \triangleq \theta_0^{n_s}-\theta_0^{n_s'} \quad (89)$$

$$\Delta\beta_{acc}(n_s,n_s',f_1,f_2) \triangleq \beta_{acc}(n_s,f_1,f_2)-\beta_{acc}(n_s',f_1,f_2) =$$
$$sgn(n_s-n_s')\times\sum_{p=\min(n_s,n_s')}^{\max(n_s,n_s')-1}\left(\beta_{2,p}\times L_s(p)+\beta_{DCU,p}\right)+$$
$$sgn(n_s-n_s')\times\pi(f_1+f_2)\sum_{p=\min(n_s,n_s')}^{\max(n_s,n_s')-1}\left(\beta_{3,p}\times L_s(p)\right)$$
$$-sgn(n_s-n_s')\times 2\pi\sum_{p=\min(n_s,n_s')}^{\max(n_s,n_s')-1}\left(\beta_{3,p}\times f_p^c\times L_s(p)\right)$$
$$sgn(x)\triangleq\begin{cases}1 & x\geq 0\\ -1 & x<0\end{cases} \quad (90)$$

Equation (88) can be written as:

$$|LK(f_1,f_2,f_3)|^2 \cong \sum_{n_s=1}^{N_s}\gamma_{n_s}^2\times g_0^2(n_s)\times\ |\xi(n_s,f_1,f_2,f)|^2+$$
$$\sum_{n_s=1}^{N_s}\sum_{n_s'=1,n_{s'}\neq n_s}^{N_s}\gamma_{n_s}\gamma_{n_s'}\times g_0(n_s')g_0(n_s)\times\ \xi(n_s,f_1,f_2,f)\times\xi^*(n_s',f_1,f_2,f) \quad (91)$$

| | |
|---|---|
| $\times\ e^{j\left[4\pi^2(f_1-f)(f_2-f)\times\Delta\beta_{acc}(n_s,n_s',f_1,f_2)+\Delta\theta(n_s,n_s')\right]}$ | |

Equation (91) can be further simplified as:

| | |
|---|---|
| $\lvert LK(f_1,f_2,f_3)\rvert^2 \cong \sum_{n_s=1}^{N_s} \gamma_{n_s}^2 \times g_0^2(n_s) \times\ \lvert\xi(n_s,f_1,f_2,f)\rvert^2 +$ $\sum_{n_s=1}^{N_s}\sum_{n_s'=n_s+1}^{N_s} \gamma_{n_s}\gamma_{n_s'} \times g_0(n_s')g_0(n_s) \times\ 2\times Real\{\xi(n_s,f_1,f_2,f)$ $\times\ \xi^*(n_s',f_1,f_2,f) \times\ e^{j\left[4\pi^2(f_1-f)(f_2-f)\times\Delta\beta_{acc}(n_s,n_s',f_1,f_2)+\Delta\theta(n_s,n_s')\right]}\}$ | (92) |

Using decomposition of $\xi$ defined in equal (86) and with the help of symbolic tools of Matlab software we will have:

| | |
|---|---|
| $\lvert\xi(n_s,f_1,f_2,f)\rvert^2 = \dfrac{J_1(n_s)}{1+{f_1'}^2{f_2'}^2D_1^2(n_s,f_1',f_2')} + \dfrac{J_2(n_s)}{1+{f_1'}^2{f_2'}^2D_2^2(n_s,f_1',f_2')}$ | (93) |

Where in (93):

| | |
|---|---|
| $f_1' \triangleq f_1 - f$ | (94) |
| $f_2' \triangleq f_2 - f$ | (95) |
| $D_1(n_s,f_1',f_2') \triangleq \dfrac{4\times\pi^2\times\left(\beta_{2,n_s}+\pi\beta_{3,n_s}\left(f_1'+f_2'+2f-2f_{n_s}^c\right)\right)}{2\overline{\alpha_0}(n_s)+\bar{\sigma}\ (n_s)}$ | (96) |
| $D_2(n_s,f_1',f_2') \triangleq \dfrac{4\times\pi^2\times\left(\beta_{2,n_s}+\pi\beta_{3,n_s}\left(f_1'+f_2'+2f-2f_{n_s}^c\right)\right)}{2\overline{\alpha_0}(n_s)}$ | (97) |
| $J_1(n_s) \triangleq \dfrac{4\overline{\alpha_1}(n_s)\times(2\overline{\alpha_0}(n_s)-\overline{\alpha_1}(n_s)+\bar{\sigma}\ (n_s))}{\bar{\sigma}\ (n_s)\times\left(2\overline{\alpha_0}(n_s)+\bar{\sigma}\ (n_s)\right)^2\times(4\overline{\alpha_0}(n_s)+\bar{\sigma}\ (n_s))}$ | (98) |
| $J_2(n_s) \triangleq \dfrac{\left(\bar{\sigma}\ (n_s)-2\overline{\alpha_1}(n_s)\right)\times(4\overline{\alpha_0}(n_s)-2\overline{\alpha_1}(n_s)+\bar{\sigma}\ (n_s))}{4\times\left(\overline{\alpha_0}(n_s)\right)^2\times\bar{\sigma}\ (n_s)\times(4\overline{\alpha_0}(n_s)+\bar{\sigma}\ (n_s))}$ | (99) |

As it is evident from (96) and (97), in general $D_1$and $D_2$ are function of $f_1'$and $f_2'$ and therefore 2-D analytic integration of expression (93) is impossible. So, here we make an approximation and instead of $D_1^2(n_s,f_1',f_2')$ and $D_2^2(n_s,f_1',f_2')$, we replace them with their averages in the integration island. We define:

| | |
|---|---|
| $$\overline{D_1}(n_s) \triangleq \pm \sqrt{\frac{\int_{f_2^*-\frac{L_2}{2}}^{f_2^*+\frac{L_2}{2}} \int_{f_1^*-\frac{L_1}{2}}^{f_1^*+\frac{L_1}{2}} \left(D_1(n_s, f_1', f_2')\right)^2 df_1 df_2}{\int_{f_2^*-\frac{L_2}{2}}^{f_2^*+\frac{L_2}{2}} \int_{f_1^*-\frac{L_1}{2}}^{f_1^*+\frac{L_1}{2}} df_1 df_2}}$$ $$= \pm \sqrt{\frac{\int_{f_2^*-f-\frac{L_2}{2}}^{f_2^*-f+\frac{L_2}{2}} \int_{f_1^*-f-\frac{L_1}{2}}^{f_1^*-f+\frac{L_1}{2}} \left(D_1(n_s, f_1', f_2')\right)^2 df_1' df_2'}{\int_{f_2^*-f-\frac{L_2}{2}}^{f_2^*-f+\frac{L_2}{2}} \int_{f_1^*-f-\frac{L_1}{2}}^{f_1^*-f+\frac{L_1}{2}} df_1' df_2'}}$$ $$= \pm \frac{4 \times \pi^2}{2\overline{\alpha_0}(n_s) + \bar{\sigma}\,(n_s)} \times$$ $$\times \sqrt{\frac{\int_{f_2^*-f-\frac{L_2}{2}}^{f_2^*-f+\frac{L_2}{2}} \int_{f_1^*-f-\frac{L_1}{2}}^{f_1^*-f+\frac{L_1}{2}} \left(\beta_{2,n_s} + \pi\beta_{3,n_s}\left(f_1' + f_2' + 2f - 2f_{n_s}^c\right)\right)^2 df_1' df_2'}{L_1 \times L_2}} =$$ $$\pm \frac{4 \times \pi^2 \sqrt{\left(\beta_{2,n_s} + \pi\beta_{3,n_s}\left(f_1^* + f_2^* - 2 \times f_{n_s}^c\right)\right)^2 + (L_1^2 + L_2^2) \times (\pi^2 \times \beta_{3,n_s}^2)/12}}{2\overline{\alpha_0}(n_s) + \bar{\sigma}\,(n_s)}$$ $$= \frac{4 \times \pi^2 \beta_{2,n_s,eff}}{2\overline{\alpha_0}(n_s) + \bar{\sigma}\,(n_s)}$$ | (100) |

Which $\beta_{2,n_s,eff}$ in (100) is defined as:

| | |
|---|---|
| $$\beta_{2,n_s,eff} \triangleq sgn(\beta_{2,n_s} + \pi\beta_{3,n_s}(f_1^* + f_2^* - 2 \times f_c)) \times$$ $$\sqrt{\left(\beta_{2,n_s} + \pi\beta_{3,n_s}\left(f_1^* + f_2^* - 2 \times f_{n_s}^c\right)\right)^2 + (L_1^2 + L_2^2) \times (\pi^2 \times \beta_{3,n_s}^2)/12}$$ $$sgn(x) \triangleq \begin{cases} 1 & x \geq 0 \\ -1 & x < 0 \end{cases}$$ | (101) |

Note that in (101) we choosed the sign of $\beta_{2,n_s,eff}$ equal to the sign of $\beta_{2,n_s} + \pi\beta_{3,n_s}\left(f_1 + f_2 - 2f_{n_s}^c\right)$ in the center of the intagration island.
Similarly for $D_2^2(n_s, f_1', f_2')$ we have:

| | |
|---|---|
| $$\overline{D_2}(n_s) \triangleq \frac{4 \times \pi^2 \beta_{2,n_s,eff}}{2\overline{\alpha_0}(n_s)}$$ | (102) |

Also in (92) we can calculate with the help of symbolic tools of Matlab :

| | |
|---|---|
| $$2Real\left\{\xi(n_s, f_1, f_2, f)\xi^*(n_s', f_1, f_2, f) \times e^{j\left[4\pi^2(f_1-f)(f_2-f)\times\Delta\beta_{acc}(n_s, n_s', f_1, f_2) + \Delta\theta(n_s, n_s')\right]}\right\}$$ $$= \sum_{p=1}^{4}\left(\frac{J_p'(n_s, n_s') \times cos(K(n_s, n_s', f_1', f_2') \times f_1' f_2' + \Delta\theta(n_s, n_s'))}{1 + {f_1'}^2 {f_2'}^2 {D_p'}^2(n_s, n_s', f_1', f_2')} + \frac{J_p''(n_s, n_s') \times f_1' f_2' \times sin(K(n_s, n_s', f_1', f_2') \times f_1' f_2' + \Delta\theta(n_s, n_s'))}{1 + {f_1'}^2 {f_2'}^2 {D_p'}^2(n_s, n_s', f_1', f_2')}\right)$$ | (103) |

Where in (103) we have :

| | |
|---|---|
| $K(n_s, n_s', f_1', f_2') \triangleq 4 \times \pi^2 \times \Delta\beta_{acc}\big(n_s, n_s', (f_1' + f), (f_2' + f)\big)$ $= 4 \times \pi^2 \times sgn(n_s - n_s') \Bigg\{ \sum_{p=\min(n_s,n_s')}^{\max(n_s,n_s')-1} \big(\beta_{2,p} \times L_s(p) + \beta_{DCU,p}\big) +$ $\pi(f_1' + f_2' + 2f) \sum_{p=\min(n_s,n_s')}^{\max(n_s,n_s')-1} \Big(\beta_{3,p}$ $\times L_s(p)\Big) - 2\pi \sum_{p=\min(n_s,n_s')}^{\max(n_s,n_s')-1} \Big(\beta_{3,p} \times f_p^c \times L_s(p)\Big) \Bigg\} =$ $4 \times \pi^2 \times \ sgn(n_s - n_s')$ $\times \sum_{p=\min(n_s,n_s')}^{\max(n_s,n_s')-1} \Big(\Big(\beta_{2,p} + \pi\beta_{3,p}\big(f_1' + f_2' + 2f - 2f_p^c\big)\Big) \times L_s(p) + \beta_{DCU,p}\Big)$ | (104) |

For notation simplicity, we define:

| | |
|---|---|
| $\beta_{2.acc}(n_s.n_{s'}) \triangleq sgn(n_s - n_s') \times \sum_{p=\min(n_s,n_s')}^{\max(n_s,n_s')-1} \big(\beta_{2,p} \times L_s(p) + \beta_{DCU,p}\big)$ | (105) |
| $\beta_{3.acc}(n_s.n_{s'}) \triangleq sgn(n_s - n_s') \times \sum_{p=\min(n_s,n_s')}^{\max(n_s,n_s')-1} \Big(\beta_{3,p} \times L_s(p)\Big)$ | (106) |

Therefore, $K(n_s, n_s', f_1', f_2')$ in (104), can be written as:

| | |
|---|---|
| $K(n_s, n_s', f_1', f_2') =$ $4 \times \pi^2 \times \Big(\beta_{2.acc}(n_s.n_{s'}) + \pi \times \beta_{3.acc}(n_s.n_{s'}) \times \big(f_1' + f_2' + 2f - 2f_p^c\big)\Big)$ | (107) |

Also in (103), $D_p'$ values are:

| | |
|---|---|
| $D_1'(n_s, n_s', f_1', f_2') \triangleq \frac{4 \times \pi^2 \times \Big(\beta_{2,n_s} + \pi\beta_{3,n_s}\big(f_1' + f_2' + 2f - 2f_{n_s}^c\big)\Big)}{2\overline{\alpha_0}(n_s) + \bar{\sigma}\ (n_s)}$ | (108) |
| $D_2'(n_s, n_s', f_1', f_2') \triangleq \frac{4 \times \pi^2 \times \Big(\beta_{2,n_s'} + \pi\beta_{3,n_s'}\big(f_1' + f_2' + 2f - 2f_{n_s}^c\big)\Big)}{2\overline{\alpha_0}(n_s') + \bar{\sigma}\ (n_s')}$ | (109) |
| $D_3'(n_s, n_s', f_1', f_2') \triangleq \frac{4\times\pi^2\times\left(\beta_{2,n_s}+\pi\beta_{3,n_s}\left(f_1'+f_2'+2f-2f_{n_s}^c\right)\right)}{2\overline{\alpha_0}(n_s)}$ | (110) |
| $D_4'(n_s, n_s', f_1', f_2') \triangleq \frac{4\times\pi^2\times\left(\beta_{2,n_s'}+\pi\beta_{3,n_s'}\left(f_1'+f_2'+2f-2f_{n_s}^c\right)\right)}{2\overline{\alpha_0}(n_s')}$ | (111) |

As we see from (108)-(111), $D_p'$ is a function of $f_1'$and $f_2'$ and therefore 2-D analytic integration of expression (100) is impossible. Similar to the approximation we performed in equation (100), we define:

| | |
|---|---|
| $\overline{D_1'}(n_s, n_s') \triangleq \frac{4\times\pi^2\times\beta_{2,n_s,eff}}{2\overline{\alpha_0}(n_s)+\bar{\sigma}\,(n_s)}$ | (112) |
| $\overline{D_2'}(n_s, n_s') \triangleq \frac{4\times\pi^2\times\beta_{2,n_s',eff}}{2\overline{\alpha_0}(n_s')+\bar{\sigma}\,(n_s')}$ | (113) |
| $\overline{D_3'}(n_s, n_s') \triangleq \frac{4\times\pi^2\times\beta_{2,n_s,eff}}{2\overline{\alpha_0}(n_s)}$ | (114) |
| $\overline{D_4'}(n_s, n_s') \triangleq \frac{4\times\pi^2\times\beta_{2,n_s',eff}}{2\overline{\alpha_0}(n_s')}$ | (115) |

Another important issue for the analytical integration of the equation (103), which is evident from (104), is that $K(n_s, n_s', f_1', f_2')$ is a function of $f_1'$and $f_2'$ and this makes the analytic integration impossible. Indeed, we must look for a $\overline{K}(n_s, n_s')$ that $K(n_s, n_s', f_1', f_2')$ can be approximately replaced by it as:

| | |
|---|---|
| $K(n_s, n_s', f_1', f_2')\times f_1'f_2'+\Delta\theta(n_s, n_s')\overset{?}{\cong}\overline{K_1}(n_s, n_s')\times f_1'f_2'+\overline{K_2}(n_s, n_s')\times f_1' + \overline{K_3}(n_s, n_s')\times f_2'+\overline{K_4}(n_s, n_s')$ | (116) |

Where in (116), $\overline{K_i}(n_s, n_s')\,; i = 1,2,3,4$ are independent of $f_1'$ and $f_2'$. Infact, if the argument of $\cos(,)$ and $\sin(,)$ in (103) are in the form of the right hand side of equation (116), the analytic integration is possible while if it is in the current form (left side of equation (116)), the analytic integration is not possible. For finding a suitable $\overline{K_i}(n_s, n_s')$, we try to force the lef and right sides of equation (116) to be in a minimum distance over the integration island. We define distance over the integration islans as:

| | |
|---|---|
| $d(\overline{K_1},\overline{K_2},\overline{K_3},\overline{K_4}) \triangleq \int_{f_2^*-f-\frac{L_2}{2}}^{f_2^*-f+\frac{L_2}{2}}\int_{f_1^*-f-\frac{L_1}{2}}^{f_1^*-f+\frac{L_1}{2}}\left(K(n_s, n_s', f_1', f_2')\times f_1'f_2'+\Delta\theta(n_s, n_s')-\overline{K_1}\times f_1'f_2' - \overline{K_2}\times f_1'-\overline{K_3}\times f_2'-\overline{K_4}\right)^2 df_1'df_2'$ | (117) |

Replacing $K(n_s, n_s', f_1', f_2')$ from equation (107) into (117), we can find $\overline{K_1},\overline{K_2},\overline{K_3},\overline{K_4}$ in such a way that $d(\overline{K_1},\overline{K_2},\overline{K_3},\overline{K_4})$ becomes minimum. For achieving this, we must have:

| | |
|---|---|
| $\frac{\partial d(\overline{K_1},\overline{K_2},\overline{K_3},\overline{K_4})}{\partial \overline{K_\iota}} = 0\,; for\ i = 1,2,3,4$ | (118) |

Therefore, taking derivative of the right side of equation (117) with respect to $\overline{K_\iota}$ and put the four equations equal to zero, we will have a set of 4 linear equations with four unknown values, $\overline{K_1},\overline{K_2},\overline{K_3},\overline{K_4}$, that can be easily solved using symbolic tools of Matlab software and finally we find :

| | |
|---|---|
| $\overline{K_1}(n_s,n_s') = 4\pi^2 \times \left(\beta_{2.acc}(n_s.n_{s'}) + 2\times\pi\times\beta_{3.acc}(n_s.n_{s'})\times\left(f_1^* + f_2^* - f - f_p^c\right)\right)$ | (119) |
| $\overline{K_2}(n_s,n_s') = \frac{\pi^3}{3}\times\beta_{3.acc}(n_s.n_{s'})\times(L_2^2 - 12\times(f_2^* - f)^2)$ | (120) |
| $\overline{K_3}(n_s,n_s') = \frac{\pi^3}{3}\times\beta_{3.acc}(n_s.n_{s'})\times(L_1^2 - 12\times(f_1^* - f)^2)$ | (121) |
| $\overline{K_4}(n_s,n_s') = \overline{\Delta\theta}(n_s,n_s')$ | (122) |

So we can rewrite equations (93) and (103) with approximations made above:

| | |
|---|---|
| $\lvert\xi(n_s,f_1,f_2,f)\rvert^2 \cong \frac{J_1(n_s)}{1+f_1'^2 f_2'^2\left(\overline{D_1}(n_s)\right)^2} + \frac{J_2(n_s)}{1+f_1'^2 f_2'^2\left(\overline{D_2}(n_s)\right)^2}$ | (123) |
| $2Real\left\{\xi(n_s,f_1,f_2,f)\xi^*(n_s',f_1,f_2,f) \times e^{j\left[4\pi^2(f_1-f)(f_2-f)\times\Delta\beta_{acc}(n_s,n_s',f_1,f_2)+\Delta\theta(n_s,n_s')\right]}\right\} \cong \sum_{p=1}^{4}\left(\frac{J_p'(n_s,n_s')}{1+f_1'^2 f_2'^2\left(\overline{D_p'}(n_s,n_s')\right)^2} \times cos\left(\overline{K_1}(n_s,n_s')\times f_1' f_2' + \overline{K_2}(n_s,n_s')\times f_1' + \overline{K_3}(n_s,n_s')\times f_2' + \overline{K_4}(n_s,n_s')\right) + \frac{J_p''(n_s,n_s')\times f_1' f_2'}{1+f_1'^2 f_2'^2\left(\overline{D_p'}(n_s,n_s')\right)^2} \times sin\left(\overline{K_1}(n_s,n_s')\times f_1' f_2' + \overline{K_2}(n_s,n_s')\times f_1' + \overline{K_3}(n_s,n_s')\times f_2' + \overline{K_4}(n_s,n_s')\right)\right)$ | (124) |

The only unknown values in equation (124) are coefficients $J_p'(n_s,n_s')$ and $J_p''(n_s,n_s')$ that arise from fraction decomposition of equation (86) and then calculating the expression

$$2Real\{\xi(n_s, f_1, f_2, f)\xi^*(n_s', f_1, f_2, f) \times e^{j[4\pi^2(f_1-f)(f_2-f)\times\Delta\beta_{acc}(n_s,n_s',f_1,f_2)+\Delta\theta(n_s,n_s')]}\}$$

For notation simplicity we first define:

| | |
|---|---|
| $B_p \triangleq 4 \times \pi^2 \times \beta_{2,p,eff}$ | (125) |

Again with the help of the Matlab symbolic tools we will have:

| | |
|---|---|
| $J_1'(n_s, n_s') \triangleq \dfrac{J_{1,N}'(n_s, n_s')}{J_{1,D}'(n_s, n_s')}$ | (126) |

Where in (126) :

| | |
|---|---|
| $J_{1,N}'(n_s, n_s') \triangleq$ $4B_{n_s}\overline{\alpha_1}(n_s) \times (2B_{n_s}\overline{\alpha_0}(n_s') + 2B_{n_s'}\overline{\alpha_0}(n_s) - 2B_{n_s}\overline{\alpha_1}(n_s') + B_{n_s}\bar{\sigma}(n_s') + B_{n_s'}\bar{\sigma}(n_s))$ | (127) |
| $J_{1,D}'(n_s, n_s') \triangleq$ $\bar{\sigma}(n_s) \times (2\overline{\alpha_0}(n_s) + \bar{\sigma}(n_s)) \times \left(2B_{n_s}\overline{\alpha_0}(n_s') + 2B_{n_s'}\overline{\alpha_0}(n_s) + B_{n_s'}\bar{\sigma}(n_s)\right) \times \left(2B_{n_s}\overline{\alpha_0}(n_s') + 2B_{n_s'}\overline{\alpha_0}(n_s) + B_{n_s}\bar{\sigma}(n_s') + B_{n_s'}\bar{\sigma}(n_s)\right)$ | (128) |
| $J_2'(n_s, n_s') \triangleq \dfrac{J_{2,N}'(n_s, n_s')}{J_{2,D}'(n_s, n_s')}$ | (129) |

Where in (129):

| | |
|---|---|
| $J_{2,N}'(n_s, n_s') \triangleq$ $4B_{n_s'}\overline{\alpha_1}(n_s') \times (2B_{n_s}\overline{\alpha_0}(n_s') + 2B_{n_s'}\overline{\alpha_0}(n_s) - 2B_{n_s'}\overline{\alpha_1}(n_s) + B_{n_s}\bar{\sigma}(n_s') + B_{n_s'}\bar{\sigma}(n_s))$ | (130) |
| $J_{2,D}'(n_s, n_s') \triangleq$ $\bar{\sigma}(n_s') \times (2\overline{\alpha_0}(n_s') + \bar{\sigma}(n_s')) \times \left(2B_{n_s}\overline{\alpha_0}(n_s') + 2B_{n_s'}\overline{\alpha_0}(n_s) + B_{n_s}\bar{\sigma}(n_s')\right) \times \left(2B_{n_s}\overline{\alpha_0}(n_s') + 2B_{n_s'}\overline{\alpha_0}(n_s) + B_{n_s}\bar{\sigma}(n_s') + B_{n_s'}\bar{\sigma}(n_s)\right)$ | (131) |
| $J_3'(n_s, n_s') \triangleq \dfrac{J_{3,N}'(n_s, n_s')}{J_{3,D}'(n_s, n_s')}$ | (132) |

Where in (132):

| | |
|---|---|
| $J_{3,N}'(n_s, n_s') \triangleq$ | (133) |
| $-B_{n_s} \times (2\overline{\alpha_1}(n_s) - \bar{\sigma}\,(n_s))$ $\times \left(2B_{n_s}\overline{\alpha_0}(n_s') + 2B_{n_s'}\overline{\alpha_0}(n_s) - 2B_{n_s}\overline{\alpha_1}(n_s') + B_{n_s}\bar{\sigma}\,(n_s')\right)$ | |
| $J_{3,D}'(n_s, n_s') \triangleq$ $2\overline{\alpha_0}(n_s)\bar{\sigma}\,(n_s) \times \left(B_{n_s}\overline{\alpha_0}(n_s') + B_{n_s'}\overline{\alpha_0}(n_s)\right)$ $\times \left(2B_{n_s}\overline{\alpha_0}(n_s') + 2B_{n_s'}\overline{\alpha_0}(n_s) + B_{n_s}\bar{\sigma}\,(n_s')\right)$ | (134) |
| $J_4'(n_s, n_s') \triangleq \dfrac{J_{4,N}'(n_s, n_s')}{J_{4,D}'(n_s, n_s')}$ | (135) |

Where in (135):

| | |
|---|---|
| $J_{4,N}'(n_s, n_s') \triangleq$ $-B_{n_s'} \times \left(2\overline{\alpha_1}(n_s') - \bar{\sigma}\,(n_s')\right)$ $\times \left(2B_{n_s}\overline{\alpha_0}(n_s') + 2B_{n_s'}\overline{\alpha_0}(n_s) - 2B_{n_s'}\overline{\alpha_1}(n_s) + B_{n_s'}\bar{\sigma}\,(n_s)\right)$ | (136) |
| $J_{4,D}'(n_s, n_s') \triangleq$ $2\overline{\alpha_0}(n_s')\bar{\sigma}\,(n_s') \times \left(B_{n_s}\overline{\alpha_0}(n_s') + B_{n_s'}\overline{\alpha_0}(n_s)\right)$ $\times \left(2B_{n_s}\overline{\alpha_0}(n_s') + 2B_{n_s'}\overline{\alpha_0}(n_s) + B_{n_s'}\bar{\sigma}\,(n_s)\right)$ | (137) |
| $J_1''(n_s, n_s') \triangleq \dfrac{J_{1,N}''(n_s, n_s')}{J_{1,D}''(n_s, n_s')}$ | (138) |

Where in (138) :

| | |
|---|---|
| $J_{1,N}''(n_s, n_s') \triangleq$ $-4B_{n_s}^2\overline{\alpha_1}(n_s) \times \left(2B_{n_s}\overline{\alpha_0}(n_s') + 2B_{n_s'}\overline{\alpha_0}(n_s) - 2B_{n_s}\overline{\alpha_1}(n_s') + B_{n_s}\bar{\sigma}\,(n_s') + B_{n_s'}\bar{\sigma}\,(n_s)\right)$ | (139) |
| $J_{1,D}''(n_s, n_s') \triangleq$ $\bar{\sigma}\,(n_s) \times \left(2\overline{\alpha_0}(n_s) + \bar{\sigma}\,(n_s)\right)^2 \times \left(2B_{n_s}\overline{\alpha_0}(n_s') + 2B_{n_s'}\overline{\alpha_0}(n_s) + B_{n_s'}\bar{\sigma}\,(n_s)\right) \times$ $\left(2B_{n_s}\overline{\alpha_0}(n_s') + 2B_{n_s'}\overline{\alpha_0}(n_s) + B_{n_s}\bar{\sigma}\,(n_s') + B_{n_s'}\bar{\sigma}\,(n_s)\right)$ | (140) |
| $J_2''(n_s, n_s') \triangleq \dfrac{J_{2,N}''(n_s, n_s')}{J_{2,D}''(n_s, n_s')}$ | (141) |

Where in (141):

| | |
|---|---|
| $J_{2,N}''(n_s, n_s') \triangleq$ | (142) |
| $4B_{n_s'}^2\overline{\alpha_1}(n_s') \times \left(2B_{n_s}\overline{\alpha_0}(n_s') + 2B_{n_s'}\overline{\alpha_0}(n_s) - 2B_{n_s'}\overline{\alpha_1}(n_s) + B_{n_s}\bar{\sigma}\,(n_s') + B_{n_s'}\bar{\sigma}\,(n_s)\right)$ | |
| $J_{2,D}''(n_s, n_s') \triangleq \bar{\sigma}\,(n_s') \times \left(2\overline{\alpha_0}(n_s') + \bar{\sigma}\,(n_s')\right)^2 \times \left(2B_{n_s}\overline{\alpha_0}(n_s') + 2B_{n_s'}\overline{\alpha_0}(n_s) + B_{n_s}\bar{\sigma}\,(n_s')\right) \times \left(2B_{n_s}\overline{\alpha_0}(n_s') + 2B_{n_s'}\overline{\alpha_0}(n_s) + B_{n_s}\bar{\sigma}\,(n_s') + B_{n_s'}\bar{\sigma}\,(n_s)\right)$ | (143) |
| $J_3''(n_s, n_s') \triangleq \frac{J_{3,N}''(n_s, n_s')}{J_{3,D}''(n_s, n_s')}$ | (144) |

Where in (144):

| | |
|---|---|
| $J_{3,N}''(n_s, n_s') \triangleq B_{n_s}^2 \times (2\overline{\alpha_1}(n_s) - \bar{\sigma}\,(n_s)) \times \left(2B_{n_s}\overline{\alpha_0}(n_s') + 2B_{n_s'}\overline{\alpha_0}(n_s) - 2B_{n_s}\overline{\alpha_1}(n_s') + B_{n_s}\bar{\sigma}\,(n_s')\right)$ <br> $J_{3,D}''(n_s, n_s') \triangleq 4\left(\overline{\alpha_0}(n_s)\right)^2 \times \bar{\sigma}\,(n_s) \times \left(B_{n_s}\overline{\alpha_0}(n_s') + B_{n_s'}\overline{\alpha_0}(n_s)\right) \times \left(2B_{n_s}\overline{\alpha_0}(n_s') + 2B_{n_s'}\overline{\alpha_0}(n_s) + B_{n_s}\bar{\sigma}\,(n_s')\right)$ | (145) |
| $J_4''(n_s, n_s') \triangleq \frac{J_{4,N}''(n_s, n_s')}{J_{4,D}''(n_s, n_s')}$ | (146) |

Where in (146):

| | |
|---|---|
| $J_{4,N}''(n_s, n_s') \triangleq -B_{n_s'}^2 \times \left(2\overline{\alpha_1}(n_s') - \bar{\sigma}\,(n_s')\right) \times \left(2B_{n_s}\overline{\alpha_0}(n_s') + 2B_{n_s'}\overline{\alpha_0}(n_s) - 2B_{n_s'}\overline{\alpha_1}(n_s) + B_{n_s'}\bar{\sigma}\,(n_s)\right)$ | (147) |
| $J_{4,D}''(n_s, n_s') \triangleq 4\left(\overline{\alpha_0}(n_s')\right)^2 \times \bar{\sigma}\,(n_s') \times \left(B_{n_s}\overline{\alpha_0}(n_s') + B_{n_s'}\overline{\alpha_0}(n_s)\right) \times \left(2B_{n_s}\overline{\alpha_0}(n_s') + 2B_{n_s'}\overline{\alpha_0}(n_s) + B_{n_s'}\bar{\sigma}\,(n_s)\right)$ | (148) |

Therefore, using equations (93) and (124), equation (92) can be written as:

| | |
|---|---|
| $$\|LK(f_1,f_2,f_3)\|^2 \cong \sum_{n_s=1}^{N_s} \gamma_{n_s}^2 \times g_0^2(n_s) \times \left( \frac{J_1(n_s)}{1+f_1'^2 f_2'^2 \left(\overline{D_1}(n_s)\right)^2} + \frac{J_2(n_s)}{1+f_1'^2 f_2'^2 \left(\overline{D_2}(n_s)\right)^2} \right) + \sum_{n_s=1}^{N_s} \sum_{n_s'=n_s+1}^{N_s} \gamma_{n_s}\gamma_{n_s'} \times g_0(n_s') g_0(n_s) \times \sum_{p=1}^{4} \left( \frac{J_p'(n_s,n_s')}{1+f_1'^2 f_2'^2 \left(\overline{D_p'}(n_s,n_s')\right)^2} \times cos\left(\overline{K_1}(n_s,n_s') \times f_1' f_2' + \overline{K_2}(n_s,n_s') \times f_1' + \overline{K_3}(n_s,n_s') \times f_2' + \overline{K_4}(n_s,n_s')\right) + \frac{J_p''(n_s,n_s') \times f_1' f_2'}{1+f_1'^2 f_2'^2 \left(\overline{D_p'}(n_s,n_s')\right)^2} \times sin\left(\overline{K_1}(n_s,n_s') \times f_1' f_2' + \overline{K_2}(n_s,n_s') \times f_1' + \overline{K_3}(n_s,n_s') \times f_2' + \overline{K_4}(n_s,n_s')\right) \right)$$ | (149) |

Also using (149), equation (54) can be written as:

| | |
|---|---|
| $$G_{NLI}(f) \cong \frac{16}{27} \sum_{m_{ch}=1}^{N_c} \sum_{n_{ch}=1}^{N_c} \sum_{k_{ch}=1}^{N_c} G_{m_{ch}} G_{n_{ch}} G_{k_{ch}} \times \int_{f_2^*-\frac{L_2}{2}}^{f_2^*+\frac{L_2}{2}} \int_{f_1^*-\frac{L_1}{2}}^{f_1^*+\frac{L_1}{2}} \left\{ \sum_{n_s=1}^{N_s} \gamma_{n_s}^2 \times g_0^2(n_s) \times \left( \frac{J_1(n_s)}{1+f_1'^2 f_2'^2 \left(\overline{D_1}(n_s)\right)^2} + \frac{J_2(n_s)}{1+f_1'^2 f_2'^2 \left(\overline{D_2}(n_s)\right)^2} \right) + \right.$$ | (150) |

$$
\sum_{n_s=1}^{N_s} \sum_{n_s'=n_s+1}^{N_s} \gamma_{n_s}\gamma_{n_s'} \times g_0(n_s')g_0(n_s)
\times \sum_{p=1}^{4} \left( \frac{J_p'(n_s,n_s')}{1 + {f_1'}^2 {f_2'}^2 \left(\overline{D_p'}(n_s,n_s')\right)^2}
\times cos\left(\overline{K_1}(n_s,n_s') \times f_1' f_2' + \overline{K_2}(n_s,n_s') \times f_1' + \overline{K_3}(n_s,n_s') \times f_2' + \overline{K_4}(n_s,n_s')\right)
+ \frac{J_p''(n_s,n_s') \times f_1' f_2'}{1 + {f_1'}^2 {f_2'}^2 \left(\overline{D_p'}(n_s,n_s')\right)^2}
\times sin\left(\overline{K_1}(n_s,n_s') \times f_1' f_2' + \overline{K_2}(n_s,n_s') \times f_1' + \overline{K_3}(n_s,n_s') \times f_2' + \overline{K_4}(n_s,n_s')\right)\right)\Bigg\} df_1 df_2
$$

And equation (150) can be more simplified as:

$$G_{NLI}(f) \cong \frac{16}{27}\sum_{m_{ch}=1}^{N_c}\sum_{n_{ch}=1}^{N_c}\sum_{k_{ch}=1}^{N_c} G_{m_{ch}}G_{n_{ch}}G_{k_{ch}}$$
$$\times\Bigg\{\sum_{n_s=1}^{N_s}\gamma_{n_s}^2 \times g_0^2(n_s) \times \sum_{p=1}^{2} J_p(n_s)$$
$$\times \int_{f_2^*-\frac{L_2}{2}}^{f_2^*+\frac{L_2}{2}}\int_{f_1^*-\frac{L_1}{2}}^{f_1^*+\frac{L_1}{2}} \frac{df_1 df_2}{1+f_1'^2 f_2'^2\left(\overline{D_p}(n_s)\right)^2}$$
$$+ \sum_{n_s=1}^{N_s}\sum_{n_s'=n_s+1}^{N_s}\gamma_{n_s}\gamma_{n_s'} \times g_0(n_s')g_0(n_s)$$
$$\times \sum_{p=1}^{4}\Bigg(J_p'(n_s,n_s')\int_{f_2^*-\frac{L_2}{2}}^{f_2^*+\frac{L_2}{2}}\int_{f_1^*-\frac{L_1}{2}}^{f_1^*+\frac{L_1}{2}}\Bigg\{\frac{1}{1+f_1'^2 f_2'^2\left(\overline{D_p'}(n_s,n_s')\right)^2}$$
$$\times \cos\left(\overline{K_1}(n_s,n_s') \times f_1' f_2' + \overline{K_2}(n_s,n_s') \times f_1' + \overline{K_3}(n_s,n_s') \times f_2'\right.$$
$$\left. + \overline{K_4}(n_s,n_s')\right)\Bigg\} df_1 df_2$$
$$+ J_p''(n_s,n_s')\int_{f_2^*-\frac{L_2}{2}}^{f_2^*+\frac{L_2}{2}}\int_{f_1^*-\frac{L_1}{2}}^{f_1^*+\frac{L_1}{2}}\Bigg\{\frac{f_1' f_2'}{1+f_1'^2 f_2'^2\left(\overline{D_p'}(n_s,n_s')\right)^2}$$
$$\times \sin\left(\overline{K_1}(n_s,n_s') \times f_1' f_2' + \overline{K_2}(n_s,n_s') \times f_1' + \overline{K_3}(n_s,n_s') \times f_2'\right.$$
$$\left. + \overline{K_4}(n_s,n_s')\right)\Bigg\} df_1 df_2 \Bigg)\Bigg\} \quad (151)$$

Using (94) and (95) with a change of variables under integral from $f_1, f_2$ to $f_1', f_2'$ we will have:

$$G_{NLI}(f) \cong \frac{16}{27} \sum_{m_{ch}=1}^{N_c} \sum_{n_{ch}=1}^{N_c} \sum_{k_{ch}=1}^{N_c} G_{m_{ch}} G_{n_{ch}} G_{k_{ch}}$$
$$\times \left\{ \sum_{n_s=1}^{N_s} \gamma_{n_s}^2 \times g_0^2(n_s) \times \sum_{p=1}^{2} J_p(n_s) \right.$$
$$\times \int_{f_2^*-f-\frac{L_2}{2}}^{f_2^*-f+\frac{L_2}{2}} \int_{f_1^*-f-\frac{L_1}{2}}^{f_1^*-f+\frac{L_1}{2}} \frac{df_1' df_2'}{1 + f_1'^2 f_2'^2 \left(\overline{D_p}(n_s)\right)^2}$$
$$+ \sum_{n_s=1}^{N_s} \sum_{n_s'=n_s+1}^{N_s} \gamma_{n_s} \gamma_{n_s'} \times g_0(n_s') g_0(n_s)$$
$$\times \sum_{p=1}^{4} \left( J_p'(n_s, n_s') \int_{f_2^*-f-\frac{L_2}{2}}^{f_2^*-f+\frac{L_2}{2}} \int_{f_1^*-f-\frac{L_1}{2}}^{f_1^*-f+\frac{L_1}{2}} \left\{ \frac{1}{1 + f_1'^2 f_2'^2 \left(\overline{D_p'}(n_s, n_s')\right)^2} \right. \right.$$
$$\times cos\left(\overline{K_1}(n_s, n_s') \times f_1' f_2' + \overline{K_2}(n_s, n_s') \times f_1' + \overline{K_3}(n_s, n_s') \times f_2'\right.$$
$$\left.\left. + \overline{K_4}(n_s, n_s')\right) \right\} df_1' df_2'$$
$$+ J_p''(n_s, n_s') \int_{f_2^*-f-\frac{L_2}{2}}^{f_2^*-f+\frac{L_2}{2}} \int_{f_1^*-f-\frac{L_1}{2}}^{f_1^*-f+\frac{L_1}{2}} \left\{ \frac{f_1' f_2'}{1 + f_1'^2 f_2'^2 \left(\overline{D_p'}(n_s, n_s')\right)^2} \right.$$
$$\times sin\left(\overline{K_1}(n_s, n_s') \times f_1' f_2' + \overline{K_2}(n_s, n_s') \times f_1' + \overline{K_3}(n_s, n_s') \times f_2'\right.$$
$$\left.\left.\left.\left. + \overline{K_4}(n_s, n_s')\right) \right\} df_1' df_2' \right) \right\} \qquad (152)$$

Considering equation (152), for achieving an analytical solution for the GN formula, we need to solve two general integral forms as:

$$I_1(D, K_1.K_2.K_3, K_4, x_1, x_2, y_1, y_2)$$
$$\triangleq \int_{y_1}^{y_2} \int_{x_1}^{x_2} \frac{cos(K_1 \times f_1' f_2' + K_2 \times f_1' + K_3 \times f_2' + K_4)}{1 + f_1'^2 f_2'^2 D^2} df_1' df_2' \qquad (153)$$

| | |
|---|---|
| $I_2(D, K_1.K_2.K_3, K_4, x_1, x_2, y_1, y_2)$ $\triangleq \int_{y_1}^{y_2}\int_{x_1}^{x_2} \frac{f_1^{'}f_2^{'} \times sin(K_1 \times f_1^{'}f_2^{'} + K_2 \times f_1^{'} + K_3 \times f_2^{'} + K_4)}{1 + f_1^{'2} f_2^{'2} D^2}\, df_1^{'}df_2^{'}$ | (154) |

With definitions in (153) and (154), equation (152) can be rewritten as:

$$G_{NLI}(f) \cong \frac{16}{27}\sum_{m_{ch}=1}^{N_c}\sum_{n_{ch}=1}^{N_c}\sum_{k_{ch}=1}^{N_c} G_{m_{ch}}G_{n_{ch}}G_{k_{ch}} \times \Bigg\{\sum_{n_s=1}^{N_s}\gamma_{n_s}^2 \times g_0^2(n_s) \times \sum_{p=1}^{2} J_p(n_s) \times I_1\left(\overline{D_p}(n_s), 0,0,0,0, \left(f_1^* - f - \frac{L_1}{2}\right), \left(f_1^* - f + \frac{L_1}{2}\right), \left(f_2^* - f - \frac{L_2}{2}\right), \left(f_2^* - f + \frac{L_2}{2}\right)\right)$$
$$+ \sum_{n_s=1}^{N_s}\sum_{n_s^{'}=n_s+1}^{N_s} \gamma_{n_s}\gamma_{n_s^{'}} \times g_0(n_s^{'})g_0(n_s) \times \sum_{p=1}^{4}\Bigg(J_p^{'}(n_s, n_s^{'}) \times I_1\left(\overline{D_p^{'}}(n_s, n_s^{'}), \overline{K_1}(n_s, n_s^{'}).\overline{K_2}(n_s, n_s^{'}), \overline{K_3}(n_s, n_s^{'}), \overline{K_4}(n_s, n_s^{'}), (f_1^* - f - \frac{L_1}{2}), (f_1^* - f + \frac{L_1}{2}), (f_2^* - f - \frac{L_2}{2}), (f_2^* - f + \frac{L_2}{2})\right)$$
$$+ J_p^{''}(n_s, n_s^{'}) \times I_2\left(\overline{D_p^{'}}(n_s, n_s^{'}), \overline{K_1}(n_s, n_s^{'}).\overline{K_2}(n_s, n_s^{'}), \overline{K_3}(n_s, n_s^{'}), \overline{K_4}(n_s, n_s^{'}), (f_1^* - f - \frac{L_1}{2}), (f_1^* - f + \frac{L_1}{2}), (f_2^* - f - \frac{L_2}{2}), (f_2^* - f + \frac{L_2}{2})\right)\Bigg)\Bigg\} \quad (155)$$

Integrals presented in (153) and (154) in the current forms, do not have known analytic solutions. So we make an approximation here as:

| | |
|---|---|
| $\frac{1}{1+x^2} \cong \sum_{i=1}^{3} H_i \times e^{-\tau_i x} \quad for\ x \geq 0$ | (156) |

We can find $H_i$ and $\tau_i$ in equation (156) through a simple fitting approach and find them in such a way that equation (156) holds with very good approximation. Our fitting result is:

| | |
|---|---|
| $[H_1, H_2, H_3] = [-76.70258992199933\,, 0.22567834335697\,, 77.47441920490010]$ | (157) |
| $[\tau_1, \tau_2, \tau_3] = [2.01946250412823\,, 0.322968123744975\,, 1.996636590604707]$ | (158) |

As we can see from figure (11), two functions in left and right side of equation (156) with constants found based on the fitting operation and presented in (157) and (158) are very well matched.

Also as function $f(x) = \frac{1}{1+x^2}$ is an even function, we can easily extend the domain of approximation to all real values of $x$ as:

$$\frac{1}{1+x^2} \cong \sum_{i=1}^{3} H_i \times e^{-\tau_i |x|} \quad for\ x \in \mathcal{R} \qquad (159)$$

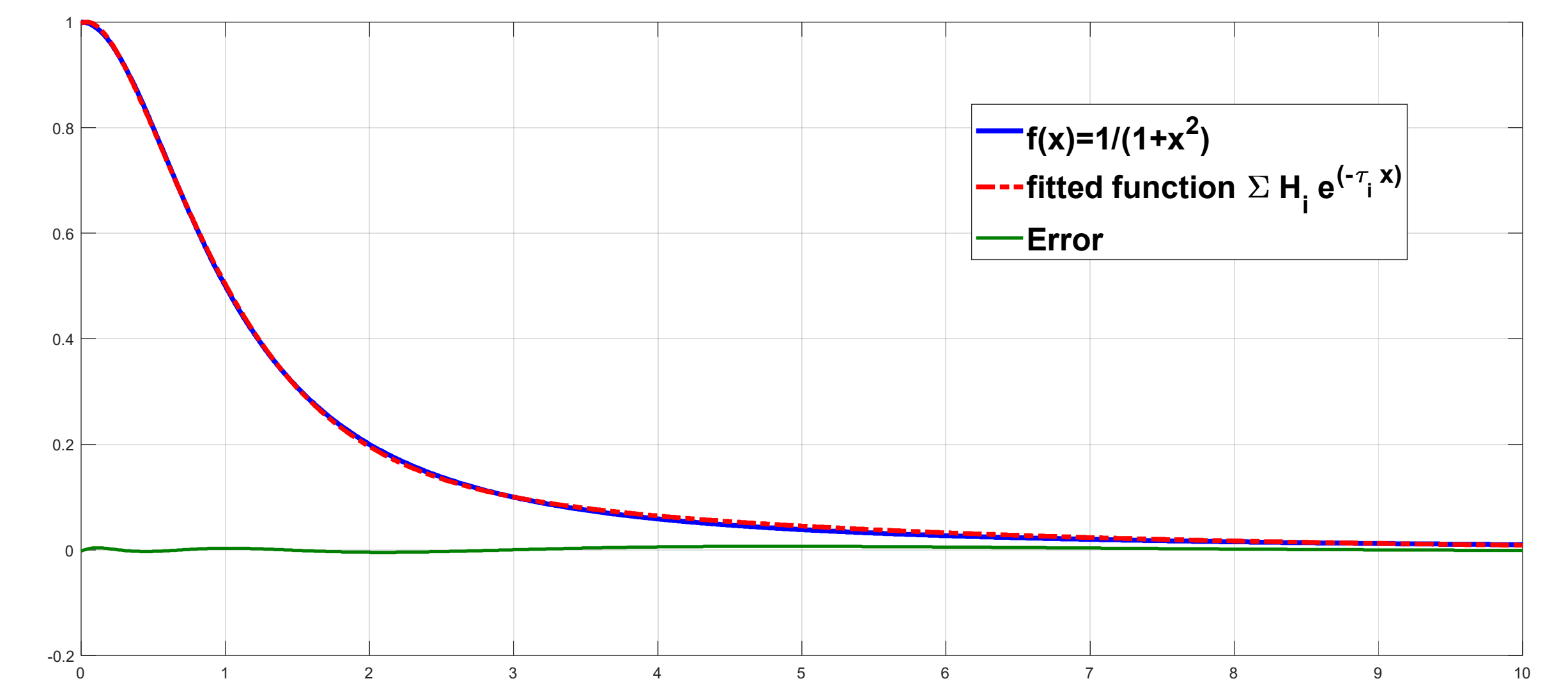

Fig(11): function $f(x) = \frac{1}{1+x^2}$ and its approximated version $\sum_{i=1}^{3} H_i \times e^{-\tau_i |x|}$

Using (159), we can write (153) and (154) in an approximate form as:

$$I_1(D, K_1.K_2.K_3, K_4, x_1, x_2, y_1, y_2) \cong \int_{y_1}^{y_2}\int_{x_1}^{x_2} \sum_{i=1}^{3} H_i \times e^{-(\tau_i \times |D|)\left|f_1^{'} f_2^{'}\right|} \times cos(K_1 \times f_1^{'} f_2^{'} + K_2 \times f_1^{'} + K_3 \times f_2^{'} + K_4)\ df_1^{'} df_2^{'} = \sum_{i=1}^{3} H_i \times \int_{y_1}^{y_2}\int_{x_1}^{x_2} e^{-(\tau_i \times |D|)\left|f_1^{'} f_2^{'}\right|} \times cos(K_1 \times f_1^{'} f_2^{'} + K_2 \times f_1^{'} + K_3 \times f_2^{'} + K_4)\ df_1^{'} df_2^{'} \qquad (160)$$

$$I_2(D, K_1.K_2.K_3, K_4, x_1, x_2, y_1, y_2) \cong \int_{y_1}^{y_2}\int_{x_1}^{x_2} \sum_{i=1}^{3} H_i \times e^{-(\tau_i \times |D|)\left|f_1^{'} f_2^{'}\right|} \times f_1^{'} f_2^{'} \times sin(K_1 \times f_1^{'} f_2^{'} + K_2 \times f_1^{'} + K_3 \times f_2^{'} + K_4)\ df_1^{'} df_2^{'} = \sum_{i=1}^{3} H_i \times \int_{y_1}^{y_2}\int_{x_1}^{x_2} e^{-(\tau_i \times |D|)\left|f_1^{'} f_2^{'}\right|} \times f_1^{'} f_2^{'} \times sin(K_1 \times f_1^{'} f_2^{'} + K_2 \times f_1^{'} + K_3 \times f_2^{'} + K_4)\ df_1^{'} df_2^{'} \qquad (161)$$

Also we define two more integral forms as below:

| | |
|---|---|
| $I_1'(B, K_1. K_2. K_3, K_4, x_1, x_2, y_1, y_2)$ $\triangleq \int_{y_1}^{y_2} \int_{x_1}^{x_2} e^{-B\left\|f_1^{'}f_2^{'}\right\|} \times cos(K_1 \times f_1^{'} f_2^{'} + K_2 \times f_1^{'} + K_3 \times f_2^{'} + K_4)\ df_1^{'} df_2^{'}$ | (162) |
| $I_2'(B, K_1. K_2. K_3, K_4, x_1, x_2, y_1, y_2)$ $\triangleq \int_{y_1}^{y_2} \int_{x_1}^{x_2} e^{-B\left\|f_1^{'}f_2^{'}\right\|} \times f_1^{'} f_2^{'} \times sin(K_1 \times f_1^{'} f_2^{'} + K_2 \times f_1^{'} + K_3 \times f_2^{'} + K_4)\ df_1^{'} df_2^{'}$ | (163) |

So having (162)-(163) we can rewrite (155) in the below form:

$$G_{NLI}(f) \cong \frac{16}{27} \sum_{m_{ch}=1}^{N_c} \sum_{n_{ch}=1}^{N_c} \sum_{k_{ch}=1}^{N_c} G_{m_{ch}} G_{n_{ch}} G_{k_{ch}}$$
$$\times \left\{ \sum_{n_s=1}^{N_s} \gamma_{n_s}^2 \times g_0^2(n_s) \times \sum_{p=1}^{2} J_p(n_s) \sum_{i=1}^{3} H_i \times \right.$$
$$\times I_1^{'} \left( \left( \tau_i \times \overline{D_p}(n_s) \right), 0,0,0,0, \left( f_1^* - f - \frac{L_1}{2} \right), \left( f_1^* - f + \frac{L_1}{2} \right), \left( f_2^* - f \right.\right.$$
$$\left.\left. - \frac{L_2}{2} \right), \left( f_2^* - f + \frac{L_2}{2} \right) \right)$$
$$+ \quad \sum_{n_s=1}^{N_s} \sum_{n_s^{'}=n_s+1}^{N_s} \gamma_{n_s} \gamma_{n_s^{'}} \times g_0(n_s^{'}) g_0(n_s)$$
$$\times \sum_{p=1}^{4} \left( J_p^{'}(n_s, n_s^{'}) \sum_{i=1}^{3} H_i \right.$$
$$\times I_1^{'} \left( \left( \tau_i \right.\right.$$
$$\left. \times \overline{D_p^{'}}(n_s, n_s^{'}) \right), \overline{K_1}(n_s, n_s^{'}). \overline{K_2}(n_s, n_s^{'}), \overline{K_3}(n_s, n_s^{'}), \overline{K_4}(n_s, n_s^{'}), (f_1^* - f$$
$$\left. - \frac{L_1}{2}), (f_1^* - f + \frac{L_1}{2}), (f_2^* - f - \frac{L_2}{2}), (f_2^* - f + \frac{L_2}{2}) \right)$$
$$+ J_p^{''}(n_s, n_s^{'}) \sum_{i=1}^{3} H_i$$
$$\times I_2^{'} \left( \left( \tau_i \right.\right.$$
$$\left. \times \overline{D_p^{'}}(n_s, n_s^{'}) \right), \overline{K_1}(n_s, n_s^{'}). \overline{K_2}(n_s, n_s^{'}), \overline{K_3}(n_s, n_s^{'}), \overline{K_4}(n_s, n_s^{'}), (f_1^* - f$$
$$\left.\left.\left. - \frac{L_1}{2}), (f_1^* - f + \frac{L_1}{2}), (f_2^* - f - \frac{L_2}{2}), (f_2^* - f + \frac{L_2}{2}) \right) \right) \right\} \quad (164)$$

For finding $I_1'$ and $I_2'$ in (162) and (163), with the help of maple software we got the 2-D indefinite integrals:

| | |
|---|---|
| $$I_1^{'',+}(B, K_1.K_2.K_3, K_4, x, y) \triangleq \iint e^{-Bxy} \times cos(K_1 \times xy + K_2 \times x + K_3 \times y + K_4)dxdy$$ $$= -Real\left\{\frac{e^{jK_4 - \frac{K_2K_3}{B-jK_1}}}{B - jK_1} \times Ei\left(\frac{(K_3 + (K_1 + jB)x)(K_2 + (K_1 + jB)y)}{B - jK_1}\right)\right\} + f_0(x) + g_0(y)$$ | (165) |
| $$I_1^{'',-}(B, K_1.K_2.K_3, K_4, x, y) \triangleq \iint e^{+Bxy} \times cos(K_1 \times xy + K_2 \times x + K_3 \times y + K_4)dxdy$$ $$= Real\left\{\frac{e^{jK_4 + \frac{K_2K_3}{B+jK_1}}}{B + jK_1} \times Ei\left(\frac{-(K_3 + (K_1 - jB)x)(K_2 + (K_1 - jB)y)}{B + jK_1}\right)\right\} f_1(x) + g_1(y)$$ | (166) |
| $$I_2^{'',+}(B, K_1.K_2.K_3, K_4, x, y) \triangleq \iint e^{-Bxy} \times xy \times \sin(K_1 \times xy + K_2 \times x + K_3 \times y + K_4)dxdy$$ $$= Real\left\{\frac{j \times (K_2K_3 + xy(B - jK_1)^2) \times e^{j(y(jBx + K_1x + K_3) + K_2x + K_4)}}{(B - jK_1)^2(K_3 + (K_1 + jB)x)(K_2 + (K_1 + jB)y)} + \frac{(B - jK_1 - K_2K_3)}{(K_1 + jB)^3} \times e^{jK_4 - \frac{K_2K_3}{B-jK_1}} \times Ei\left(\frac{(K_3 + (K_1 + jB)x)(K_2 + (K_1 + jB)y)}{B - jK_1}\right)\right\} + f_2(x) + g_2(y)$$ | (167) |
| | |
| $$I_2^{\prime\prime,-}(B,K_1.K_2.K_3,K_4,x,y) \triangleq \iint e^{+Bxy} \times xy \times \sin(K_1 \times xy + K_2 \times x + K_3 \times y + K_4)dxdy = Real\left\{\frac{j \times (K_2K_3 + xy(B+jK_1)^2) \times e^{j(y(-jBx+K_1x+K_3)+K_2x+K_4)}}{(B+jK_1)^2(K_3+(K_1-jB)x)(K_2+(K_1-jB)y)} + \frac{j(B+jK_1+K_2K_3)}{(B+jK_1)^3} \times e^{jK_4+\frac{K_2K_3}{B+jK_1}} \times Ei\left(\frac{-(K_3+(K_1-jB)x)(K_2+(K_1-jB)y)}{B+jK_1}\right)\right\} + f_3(x) + g_3(y)$$ | (168) |

Where in (164)-(168), $f_0, g_0, f_1, g_1, f_2, g_2, f_3, g_3$ are arbitary functions. Also, the function $Ei(.)$is the complex valued expotential integral which is defined as:

| | |
|---|---|
| $$Ei(z) = -\int_{-z}^{\infty} \frac{e^{-t}}{t}\,dt \quad ; \quad \lvert\arg(z)\rvert < \pi$$ | (169) |

There is a closely related function to $Ei(.)$ which is called the Theis well function. It is denoted by $E_1(.)$ and is available in Matlab software ($E_1(z) = expint(z)$) and we have[4]:

| | |
|---|---|
| $$Ei(z) = -E_1(-z) - sgn(\arg(-z))j\pi$$ $$sgn(x) \triangleq \begin{cases} -1 & x < 0 \\ 0 & x = 0 \\ 1 & x > 0 \end{cases}$$ | (170) |

We also note that $Ei(.)$ function can be represented in series representation as [7]:

| | |
|---|---|
| $$Ei(z) = \gamma + \ln(z) - j\pi\left\lfloor\frac{\arg(z)+\pi}{2\pi}\right\rfloor + \sum_{k=1}^{\infty}\frac{z^k}{k \times k!}$$ | (171) |

In our current application, we always have $\left\lfloor\frac{\arg(z)+\pi}{2\pi}\right\rfloor = 0$. Also $\gamma$ is Euler-Mascheroni constant [7] which is $\gamma \cong 0.577215665$.

By using (171), and letting $I_1^{\prime\prime,+}, I_1^{\prime\prime,-}, I_2^{\prime\prime,+}, I_2^{\prime\prime,-}$ tend to zero when $xy \to 0$ we can easily find:

| | |
|---|---|
| $$f_0(x)+g_0(y)=+Real\left\{\frac{e^{jK_4-\frac{K_2K_3}{B-jK_1}}}{B-jK_1}\times\left[Ei\left(\frac{K_3(K_2+(K_1+jB)y)}{B-jK_1}\right)+Ei\left(\frac{K_2(K_3+(K_1+jB)x)}{B-jK_1}\right)-Ei\left(\frac{K_2K_3}{B-jK_1}\right)\right]\right\}$$ | (172) |
| $$f_1(x)+g_1(y)=-Real\left\{\frac{e^{jK_4+\frac{K_2K_3}{B+jK_1}}}{B+jK_1}\times\left[Ei\left(\frac{-K_3\times(K_2+(K_1-jB)y)}{B+jK_1}\right)+Ei\left(\frac{-K_2\times(K_3+(K_1-jB)x)}{B+jK_1}\right)-Ei\left(\frac{-K_2\times K_3}{B+jK_1}\right)\right]\right\}$$ | (173) |
| $$f_2(x)+g_2(y)=-Real\left\{\frac{j\times K_2\times e^{j(yK_3+K_4)}}{(B-jK_1)^2(K_2+(K_1+jB)y)}+\frac{(B-jK_1-K_2K_3)}{(K_1+jB)^3}\times e^{jK_4-\frac{K_2K_3}{B-jK_1}}\times Ei\left(\frac{K_3(K_2+(K_1+jB)y)}{B-jK_1}\right)+\frac{j\times K_3\times e^{j(xK_2+K_4)}}{(B-jK_1)^2(K_3+(K_1+jB)x)}+\frac{(B-jK_1-K_2K_3)}{(K_1+jB)^3}\times e^{jK_4-\frac{K_2K_3}{B-jK_1}}\times Ei\left(\frac{K_2(K_3+(K_1+jB)x)}{B-jK_1}\right)-\frac{j\times e^{jK_4}}{(B-jK_1)^2}-\frac{(B-jK_1-K_2K_3)}{(K_1+jB)^3}\times e^{jK_4-\frac{K_2K_3}{B-jK_1}}\times Ei\left(\frac{K_2K_3}{B-jK_1}\right)\right\}$$ | (174) |
| $$f_3(x) + g_3(y) = -Real\left\{\frac{j \times K_2 \times e^{j(yK_3+K_4)}}{(B+jK_1)^2(K_2+(K_1-jB)y)} + \frac{j(B+jK_1+K_2K_3)}{(B+jK_1)^3} \times e^{jK_4+\frac{K_2K_3}{B+jK_1}} \times Ei\left(\frac{-K_3(K_2+(K_1-jB)y)}{B+jK_1}\right) + \frac{j \times K_3 \times e^{j(xK_2+K_4)}}{(B+jK_1)^2(K_3+(K_1-jB)x)} + \frac{j(B+jK_1+K_2K_3)}{(B+jK_1)^3} \times e^{jK_4+\frac{K_2K_3}{B+jK_1}} \times Ei\left(\frac{-K_2(K_3+(K_1-jB)x)}{B+jK_1}\right) - \frac{j \times e^{jK_4}}{(B+jK_1)^2} - \frac{j(B+jK_1+K_2K_3)}{(B+jK_1)^3} \times e^{jK_4+\frac{K_2K_3}{B+jK_1}} \times Ei\left(\frac{-K_2K_3}{B+jK_1}\right)\right\}$$ | (175) |

Therefore, replacing (171)-(175) in (165)-(168) we will have:

| | |
|---|---|
| $$I_1''^{,+}(B, K_1.K_2.K_3, K_4, x, y) = -Real\left\{\frac{e^{jK_4-\frac{K_2K_3}{B-jK_1}}}{B-jK_1} h_1(B.K_1.K_2.K_3.x.y)\right\}$$ | (176) |
| $$I_1''^{,-}(B, K_1.K_2.K_3, K_4, x, y) = Real\left\{\frac{e^{jK_4+\frac{K_2K_3}{B+jK_1}}}{B+jK_1} \times h_2(B.K_1.K_2.K_3.x.y)\right\}$$ | (177) |
| $$I_2''^{,+}(B, K_1.K_2.K_3, K_4, x, y) = Real\left\{\frac{j \times (K_2K_3 + xy(B-jK_1)^2) \times e^{j(y(jBx+\ {}_1x+K_3)+K_2x+K_4)}}{(B-jK_1)^2(K_3+(K_1+jB)x)(K_2+(K_1+jB)y)} - \frac{j \times K_2 \times e^{j(yK_3+K_4)}}{(B-jK_1)^2(K_2+(K_1+jB)y)} - \frac{j \times K_3 \times e^{j(xK_2+K_4)}}{(B-jK_1)^2(K_3+(K_1+jB)x)} + \frac{j \times e^{jK_4}}{(B-jK_1)^2} + \frac{(B-jK_1-K_2K_3)}{(K_1+jB)^3} \times e^{jK_4-\frac{K_2K_3}{B-jK_1}} \times h_1(B.K_1.K_2.K_3.x.y)\right\}$$ | (178) |

$$I_2^{'',-}(B, K_1.K_2.K_3, K_4, x, y) = Real\left\{\frac{j \times (K_2K_3 + xy(B + jK_1)^2) \times e^{j(y(-jBx+K_1x+K_3)+K_2x+K_4)}}{(B + jK_1)^2(K_3 + (K_1 - jB)x)(K_2 + (K_1 - jB)y)} - \frac{j \times K_2 \times e^{j(yK_3+K_4)}}{(B + jK_1)^2(K_2 + (K_1 - jB)y)} - \frac{j \times K_3 \times e^{j(xK_2+K_4)}}{(B + jK_1)^2(K_3 + (K_1 - jB)x)} + \frac{j \times e^{jK_4}}{(B + jK_1)^2} + \frac{j(B + jK_1 + K_2K_3)}{(B + jK_1)^3} \times e^{jK_4 + \frac{K_2K_3}{B+jK_1}} \times h_2(B.K_1.K_2.K_3.x.y)\right\} \quad (179)$$

Where $h_1(B.K_1.K_2.K_3.x.y)$ and $h_2(B.K_1.K_2.K_3.x.y)$ in (176)-(179) are:

$$h_1(B.K_1.K_2.K_3.x.y) = \left[Ei\left(\frac{(K_3 + (K_1 + jB)x)(K_2 + (K_1 + jB)y)}{B - jK_1}\right) - Ei\left(\frac{K_3(K_2 + (K_1 + jB)y)}{B - jK_1}\right) - Ei\left(\frac{K_2(K_3 + (K_1 + jB)x)}{B - jK_1}\right) + Ei\left(\frac{K_2K_3}{B - jK_1}\right)\right] \quad (180)$$

$$h_2(B.K_1.K_2.K_3.x.y) = \left[Ei\left(\frac{-(K_3 + (K_1 - jB)x)(K_2 + (K_1 - jB)y)}{B + jK_1}\right) - Ei\left(\frac{-K_3(K_2 + (K_1 - jB)y)}{B + jK_1}\right) - Ei\left(\frac{-K_2(K_3 + (K_1 - jB)x)}{B + jK_1}\right) + Ei\left(\frac{-K_2K_3}{B + jK_1}\right)\right] \quad (181)$$

We can rewrite (180) and (181) as:

$$h_1(B.K_1.K_2.K_3.x.y) = \left[Ei\left(\frac{(K_3+(K_1+jB)x)(K_2+(K_1+jB)y)}{B-jK_1}\right) - ln\left(\frac{(K_3+(K_1+jB)x)(K_2+(K_1+jB)y)}{B-jK_1}\right) - \gamma - Ei\left(\frac{K_3(K_2+(K_1+jB)y)}{B-jK_1}\right) + ln\left(\frac{K_3(K_2+(K_1+jB)y)}{B-jK_1}\right) + \gamma - Ei\left(\frac{K_2(K_3+(K_1+jB)x)}{B-jK_1}\right) + ln\left(\frac{K_2(K_3+(K_1+jB)x)}{B-jK_1}\right) + \gamma + Ei\left(\frac{K_2K_3}{B-jK_1}\right) - ln\left(\frac{K_2K_3}{B-jK_1}\right) - \gamma\right] \tag{182}$$

$$h_2(B.K_1.K_2.K_3.x.y) = \left[Ei\left(\frac{-(K_3+(K_1-jB)x)(K_2+(K_1-jB)y)}{B+jK_1}\right) - ln\left(\frac{-(K_3+(K_1-jB)x)(K_2+(K_1-jB)y)}{B+jK_1}\right) - \gamma - Ei\left(\frac{-K_3(K_2+(K_1-jB)y)}{B+jK_1}\right) + ln\left(\frac{-K_3(K_2+(K_1-jB)y)}{B+jK_1}\right) + \gamma - Ei\left(\frac{-K_2(K_3+(K_1-jB)x)}{B+jK_1}\right) + ln\left(\frac{-K_2(K_3+(K_1-jB)x)}{B+jK_1}\right) + \gamma + Ei\left(\frac{-K_2K_3}{B+jK_1}\right) - ln\left(\frac{-K_2K_3}{B+jK_1}\right) - \gamma\right] \tag{183}$$

Also using (171), we can write:

| | |
|---|---|
| $h_1(B.K_1.K_2.K_3.x.y) = \sum_{k=1}^{\infty}\frac{\left(\frac{(K_3+(K_1+jB)x)(K_2+(K_1+jB)y)}{B-jK_1}\right)^k}{k\times k!} - \sum_{k=1}^{\infty}\frac{\left(\frac{K_3(K_2+(K_1+jB)y)}{B-jK_1}\right)^k}{k\times k!} - \sum_{k=1}^{\infty}\frac{\left(\frac{K_2(K_3+(K_1+jB)x)}{B-jK_1}\right)^k}{k\times k!} + \sum_{k=1}^{\infty}\frac{\left(\frac{K_2K_3}{B-jK_1}\right)^k}{k\times k!}$ | (184) |
| $h_2(B.K_1.K_2.K_3.x.y) = \sum_{k=1}^{\infty}\frac{\left(\frac{-(K_3+(K_1-jB)x)(K_2+(K_1-jB)y)}{B+jK_1}\right)^k}{k\times k!} - \sum_{k=1}^{\infty}\frac{\left(\frac{-K_3(K_2+(K_1-jB)y)}{B+jK_1}\right)^k}{k\times k!} - \sum_{k=1}^{\infty}\frac{\left(\frac{-K_2(K_3+(K_1-jB)x)}{B+jK_1}\right)^k}{k\times k!} + \sum_{k=1}^{\infty}\frac{\left(\frac{-K_2K_3}{B+jK_1}\right)^k}{k\times k!}$ | (185) |

Note that for using (182)-(183) in a practical numerical calculation, we have singularitis for both $Ei(.)$ and $ln(.)$ functions when the absolute value of their arguments goes near zero. For avoiding this issue, we define:

| | |
|---|---|
| $h_1^{(i)}(B.K_1.K_2.K_3.x.y) \cong \begin{cases} Ei\left(z_1^{(i)}\right) - ln\left(z_1^{(i)}\right) - \gamma & \left\lvert z_1^{(i)}\right\rvert \geq Th \\ \sum_{k=1}^{N_L}\frac{\left(z_1^{(i)}\right)^k}{k\times k!} & \left\lvert z_1^{(i)}\right\rvert < Th \end{cases}; \quad for\ i = 1.2..3.4$ | (186) |
| $h_2^{(i)}(B.K_1.K_2.K_3.x.y) \cong \begin{cases} Ei\left(z_2^{(i)}\right) - ln\left(z_2^{(i)}\right) - \gamma & \left\lvert z_2^{(i)}\right\rvert \geq Th \\ \sum_{k=1}^{N_L}\frac{\left(z_2^{(i)}\right)^k}{k\times k!} & \left\lvert z_2^{(i)}\right\rvert < Th \end{cases}; \quad for\ i = 1.2..3.4$ | (187) |

Where $z_1^{(i)}$ and $z_2^{(i)}$ in (186)-(187) are:

| | |
|---|---|
| $$z_1^{(1)} = \frac{(K_3 + (K_1 + jB)x)(K_2 + (K_1 + jB)y)}{B - jK_1}$$ | (188) |
| $$z_1^{(2)} = \left(\frac{K_3(K_2 + (K_1 + jB)y)}{B - jK_1}\right)$$ | (189) |
| $$z_1^{(3)} = \left(\frac{K_2(K_3 + (K_1 + jB)x)}{B - jK_1}\right)$$ | (190) |
| $$z_1^{(4)} = \left(\frac{K_2K_3}{B - jK_1}\right)$$ | (191) |
| $$z_2^{(1)} = \left(\frac{-(K_3 + (K_1 - jB)x)(K_2 + (K_1 - jB)y)}{B + jK_1}\right)$$ | (192) |
| $$z_2^{(2)} = \left(\frac{-K_3(K_2 + (K_1 - jB)y)}{B + jK_1}\right)$$ | (193) |
| $$z_2^{(3)} = \left(\frac{-K_2(K_3 + (K_1 - jB)x)}{B + jK_1}\right)$$ | (194) |
| $$z_2^{(4)} = \left(\frac{-K_2K_3}{B + jK_1}\right)$$ | (195) |

We can set e threshold value ,i.e. $Th = 0{,}001$, and also set a number for the truncation of the infinite series , i.e. $N_L = 4$, and use (186) and (187) for avoiding any numerical ambiguity.
Therefore, we can write (176)-(179) as:

| | |
|---|---|
| $$I_1^{'',+}(B, K_1.K_2.K_3, K_4, x, y) = -Real\left\{\frac{e^{jK_4 - \frac{K_2K_3}{B - jK_1}}}{B - jK_1}\left[h_1^{(1)}(B.K_1.K_2.K_3.x.y) - h_1^{(2)}(B.K_1.K_2.K_3.x.y) - h_1^{(3)}(B.K_1.K_2.K_3.x.y) + h_1^{(4)}(B.K_1.K_2.K_3.x.y)\right]\right\}$$ | (196) |
| $I_1^{\prime\prime,-}(B,K_1.K_2.K_3,K_4,x,y)$ $= Real\left\{\frac{e^{jK_4+\frac{K_2K_3}{B+jK_1}}}{B+jK_1}\right.$ $\times\left[h_2^{(1)}(B.K_1.K_2.K_3.x.y) - h_2^{(2)}(B.K_1.K_2.K_3.x.y)\right.$ $\left.\left.- h_2^{(3)}(B.K_1.K_2.K_3.x.y) + h_2^{(4)}(B.K_1.K_2.K_3.x.y)\right]\right\}$ | (197) |
| $I_2^{\prime\prime,+}(B,K_1.K_2.K_3,K_4,x,y)$ $= Real\left\{\frac{j\times(K_2K_3+xy(B-jK_1)^2)\times e^{j(y(jBx \quad {}_1x+K_3)+K_2x+K_4)}}{(B-jK_1)^2(K_3+(K_1+jB)x)(K_2+(K_1+jB)y)}\right.$ $-\frac{j\times K_2\times e^{j(yK_3+K_4)}}{(B-jK_1)^2(K_2+(K_1+jB)y)}-\frac{j\times K_3\times e^{j(xK_2+K_4)}}{(B-jK_1)^2(K_3+(K_1+jB)x)}$ $+\frac{j\times e^{jK_4}}{(B-jK_1)^2}$ $+\frac{(B-jK_1-K_2K_3)}{(K_1+jB)^3}\times e^{jK_4-\frac{K_2K_3}{B-jK_1}}$ $\times\left[h_1^{(1)}(B.K_1.K_2.K_3.x.y) - h_1^{(2)}(B.K_1.K_2.K_3.x.y)\right.$ $\left.\left.- h_1^{(3)}(B.K_1.K_2.K_3.x.y) + h_1^{(4)}(B.K_1.K_2.K_3.x.y)\right]\right\}$ | (198) |
| $I_2^{'',-}(B,K_1.K_2.K_3,K_4,x,y)$ $= Real\left\{\frac{j\times(K_2K_3+xy(B+jK_1)^2)\times e^{j(y(-jBx+K_1x+K_3)+K_2x+K_4)}}{(B+jK_1)^2(K_3+(K_1-jB)x)(K_2+(K_1-jB)y)}\right.$ $-\frac{j\times K_2\times e^{j(yK_3+K_4)}}{(B+jK_1)^2(K_2+(K_1-jB)y)}-\frac{j\times K_3\times e^{j(xK_2+K_4)}}{(B+jK_1)^2(K_3+(K_1-jB)x)}$ $+\frac{j\times e^{jK_4}}{(B+jK_1)^2}$ $+\frac{j(B+jK_1+K_2K_3)}{(B+jK_1)^3}\times e^{jK_4+\frac{K_2K_3}{B+jK_1}}$ $\times\Big[h_2^{(1)}(B.K_1.K_2.K_3.x.y)-h_2^{(2)}(B.K_1.K_2.K_3.x.y)$ $\left.-h_2^{(3)}(B.K_1.K_2.K_3.x.y)+h_2^{(4)}(B.K_1.K_2.K_3.x.y)\Big]\right\}$ | (199) |

And also we have:

| | |
|---|---|
| $I_1''(B,K_1.K_2.K_3,K_4,x,y)\triangleq\begin{cases}I_1^{'',+}(B,K_1.K_2.K_3,K_4,x,y) & if\ (xy)\geq 0\\ I_1^{'',-}(B,K_1.K_2.K_3,K_4,x,y) & if\ (xy)<0\end{cases}$ | (200) |
| $I_2''(B,K_1.K_2.K_3,K_4,x,y)\triangleq\begin{cases}I_2^{'',+}(B,K_1.K_2.K_3,K_4,x,y) & if\ (xy)\geq 0\\ I_2^{'',-}(B,K_1.K_2.K_3,K_4,x,y) & if\ (xy)<0\end{cases}$ | (201) |

Therefore, having (186)-(201), we can write $I_1'$ and $I_2'$ in (162) - (163) as:

| | |
|---|---|
| $I_1'(B,K_1.K_2.K_3,K_4,x_1,x_2,y_1,y_2)$ $\triangleq\int_{y_1}^{y_2}\int_{x_1}^{x_2}e^{-B\lvert f_1'f_2'\rvert}\times cos\left(K_1\times f_1'f_2'+K_2\times f_1'+K_3\times f_2'+K_4\right)\,df_1'df_2'$ $=I_1''(B,K_1.K_2.K_3,K_4,x_2,y_2)+I_1''(B,K_1.K_2.K_3,K_4,x_1,y_1)$ $-I_1''(B,K_1.K_2.K_3,K_4,x_1,y_2)-I_1''(B,K_1.K_2.K_3,K_4,x_2,y_1)$ | (202) |
| $I_2'(B,K_1.K_2.K_3,K_4,x_1,x_2,y_1,y_2)$ $\triangleq\int_{y_1}^{y_2}\int_{x_1}^{x_2}e^{-B\lvert f_1'f_2'\rvert}\times f_1'f_2'\times sin\left(K_1\times f_1'f_2'+K_2\times f_1'+K_3\times f_2'+K_4\right)\,df_1'df_2'$ $=I_2''(B,K_1.K_2.K_3,K_4,x_2,y_2)+I_2''(B,K_1.K_2.K_3,K_4,x_1,y_1)$ $-I_2''(B,K_1.K_2.K_3,K_4,x_1,y_2)-I_2''(B,K_1.K_2.K_3,K_4,x_2,y_1)$ | (203) |

And finally, considering equation (164), we can find the GN model result as equation (204) which the terms sorounded by red ellipsoide are each span incoherent contribution on the total NLI and the terms sorrounded by blue ellipsoide are due to the coherent interaction of the NLI of the spans in the final NLI.

Noncoherent terms

$$
\begin{aligned}
G_{NLI}(f) \cong\ & \frac{16}{27} \sum_{m_{ch}=1}^{N_c} \sum_{n_{ch}=1}^{N_c} \sum_{k_{ch}=1}^{N_c} G_{m_{ch}} G_{n_{ch}} G_{k_{ch}} \\
& \times \left\{ \sum_{n_s=1}^{N_s} \gamma_{n_s}^2 \times g_0^2(n_s) \times \sum_{p=1}^{2} J_p(n_s) \sum_{i=1}^{3} H_i \times \right. \\
& \times I_1'\left(\left(\tau_i \times \overline{D_p}(n_s)\right), 0,0,0.0, \left(f_1^* - f - \frac{L_1}{2}\right), \left(f_1^* - f + \frac{L_1}{2}\right), \left(f_2^* - f \right.\right. \\
& \left.\left. - \frac{L_2}{2}\right), \left(f_2^* - f + \frac{L_2}{2}\right)\right) \\
& + \sum_{n_s=1}^{N_s} \sum_{n_s'=n_s+1}^{N_s} \gamma_{n_s} \gamma_{n_s'} \times g_0(n_s') g_0(n_s) \\
& \times \sum_{p=1}^{4} \left( J_p'(n_s, n_s') \sum_{i=1}^{3} H_i \right. \\
& \times I_1'\left(\left(\tau_i \times \overline{D_p'}(n_s, n_s')\right), \overline{K_1}(n_s, n_s'), \overline{K_2}(n_s, n_s'), \overline{K_3}(n_s, n_s'), \overline{K_4}(n_s, n_s'), (f_1^* - f \right. \\
& \left. - \frac{L_1}{2}), (f_1^* - f + \frac{L_1}{2}), (f_2^* - f - \frac{L_2}{2}), (f_2^* - f + \frac{L_2}{2})\right) \\
& + J_p''(n_s, n_s') \sum_{i=1}^{3} H_i \\
& \times I_2'\left(\left(\tau_i \times \overline{D_p'}(n_s, n_s')\right), \overline{K_1}(n_s, n_s'), \overline{K_2}(n_s, n_s'), \overline{K_3}(n_s, n_s'), \overline{K_4}(n_s, n_s'), (f_1^* - f \right. \\
& \left.\left.\left. - \frac{L_1}{2}), (f_1^* - f + \frac{L_1}{2}), (f_2^* - f - \frac{L_2}{2}), (f_2^* - f + \frac{L_2}{2})\right)\right)\right\}
\end{aligned}
\qquad (204)
$$

Coherent Terms

## 5. Conclusion

In this work we have presented a closed-form formula for the nonlinearity assessment in coherent fiber optic links based on the GN model. Unlike similar previous results which ignore some MCI integration islands in the 2-dimentional integration of the GN formula, we considered all possible integration islands. Also we do not make the incoherent GN approximation and considered all terms which arise due to the coherent interaction of the nonlinearity made by each fiber span. This new formula considerably extends the envelope of accurate applicability of previous results at those systems that contain one or more spans consisting of low-dispersion fiber, or even fiber whose dispersion zero falls in-band.

As a result, we believe this newly derived formula may be a very effective and general tool for supporting physical-layer aware real-time management and control of broadband C+L-band optical WDM backbone networks.

## 6. Acknowledgements

This work was sponsored by CISCO Photonics under the SRA agreement “SMART-LINKS” with Politecnico di Torino, and by the PhotoNext Center of Politecnico di Torino. The authors would like to thank Fabrizio Forghieri and Stefano Piciaccia from CISCO Photonics for their keen guidance and advice.

## 7. Appendix

In this section, we summarize the way of the calculation of the derived formula mainly in practical implementation point of view. The final formula is presented in equation (183). First of all, we have three loops through $m_{ch}$, $n_{ch}$ and $k_{ch}$ for each round of loops we need to do based on the steps below:

1- Having $m_{ch}, n_{ch}$ and $k_{ch}$ by using equations (18)-(29) and (34)-(52), we find $f_1^*$, $f_1^*$, $L_1$ and $L_2$. If $L_1$ or $L_2$ is zero, the contribution of the integration for the current $m_{ch}, n_{ch}$ and $k_{ch}$ is zero and we go to the next round of the loop and start again from beginning of step1. Otherwise, when $L_1 > 0$ and $L_2 > 0$ having calculated the numerical values of $f_1^*$, $f_1^*$, $L_1$ and $L_2$, we go to step2.

2- Using (69) and (70), we find $\overline{\alpha_{0,n_s}}$, $\overline{\alpha_{1,n_s}}$ and $\overline{\sigma_{n_s}}$ for each span.

3- Using (80), we calculate $g_0(n_s)$ for each span.

4- Based on equation (83) we calculate $\theta_0^{n_s}$ for each span and then using (89) we calculate $\Delta\theta(n_s, n_s')$ for all possible $(n_s, n_s')$.

5- Based on equation (101), we calculate $\beta_{2,n_s,eff}$ for each span.

6- Using (100) and (102), we calculate $\overline{D_1}(n_s)$ and $\overline{D_2}(n_s)$ for each span. Also using (112)-(115), we calculate $\overline{D_1'}(n_s, n_s')$, $\overline{D_2'}(n_s, n_s')$ , $\overline{D_3'}(n_s, n_s')$ and $\overline{D_4'}(n_s, n_s')$ for all possible $(n_s, n_s')$.

7- Using (119)-(122), we calculate $\overline{K_1}(n_s, n_s')$, $\overline{K_2}(n_s, n_s')$, $\overline{K_3}(n_s, n_s')$ and $\overline{K_4}(n_s, n_s')$ for all possible $(n_s, n_s')$.

8- Using (98)-(99) we calculate $J_1(n_s)$ and $J_2(n_s)$ for all spans.

9- Using (125)-(148) we find $J_1'(n_s, n_s')$, $J_2'(n_s, n_s')$ , $J_3'(n_s, n_s')$, $J_4'(n_s, n_s')$, $J_1''(n_s, n_s')$ , $J_2''(n_s, n_s')$ , $J_3''(n_s, n_s')$ and $J_4''(n_s, n_s')$ for all possible $(n_s, n_s')$.

10- $H_i$ and $\tau_i$ numerical values are presented in (157)-(158).

11- Using (186)-(203), replacing all needed parameters calculated in the previous steps, we can calculate all contributions of $I_1'()$ and $I_2'()$ in (204). Therefore, the terms circled by green and red ellipsoids can be calculated.

12- We go to the next round of our loop for the next values of $m_{ch}$, $n_{ch}$ and $k_{ch}$ and start again from step1.